\documentclass[ACS,STIX1COL]{WileyNJD-v2}

\articletype{Article Type}%

\received{26 April 2016}
\revised{6 June 2016}
\accepted{6 June 2016}

\usepackage{subfigure}
\usepackage{bm}
\usepackage{tikz}
\usetikzlibrary{shapes,shadows,arrows,positioning,patterns,shadows.blur,math}

\usepackage{soul}
   
\begin{document}

\title{Influence of weak electrostatic charges and secondary flows on pneumatic powder transport}

\author[label1,label2]{Holger Grosshans*}

\address[label1]{\orgname{Physikalisch-Technische Bundesanstalt (PTB)}, \orgaddress{Braunschweig, Germany}}
\address[label2]{\orgname{Otto von Guericke University of Magdeburg, Institute of Apparatus- and Environmental Technology}, \orgaddress{Magdeburg, Germany}}

\authormark{H. Grosshans}

\corres{*H. Grosshans. \email{holger.grosshans@ptb.de}}

\abstract[Summary]{
The dynamics of the transported powder determines the functionality and safety of pneumatic conveying systems.
The relation between the carrier gas flow, induced by the flown-through geometry, and the powder flow pattern is not clear yet for electrostatically charged particles.
This paper highlights the influence of relatively minor cross-sectional secondary flows and electrostatic forces on the concentration and dynamics of the particles.
To this end, direct numerical simulations (DNS) capture the interaction of the continuous and dispersed phases by a four-way coupled Eulerian-Lagrangian strategy.
The transport of weakly charged particles in channel flows, where turbopheresis defines the particle concentration, is compared to duct flows, where additional cross-sectional vortices form.
For both geometries the Stokes number ($S\!t=8, 32$) and the electrical Stokes number ($S\!t_\mathrm{el}=[0, 1, 2, 4]\times 10^{-3}$) are varied, the turbulent carrier flow was fixed to $Re_\tau=360$.
The presented simulations demonstrate that secondary flows, for the same $Re_\tau$, $S\!t$, and $S\!t_\mathrm{el}$, dampen the effect of particle charge.
In a duct flow, vortical secondary flows enhance the cross-sectional particle mobility against the direction of electrostatic forces.
Compared to a duct flow, in a channel the wall-normal aerodynamic forces are weaker.
Thus, electrical forces dominate their transport; the local particle concentration at the walls increases.
Further, electrostatic charges cause a stronger correlation between the gas and particle velocities.
In conclusion, despite being weak compared to the primary flow forces, secondary flow and electrostatic forces drive particle dynamics during pneumatic transport.
}

\keywords{
direct numerical simulation, particle-laden flows, electrostatics, secondary flows, channel, duct
}


\maketitle


\section{Introduction}

Particle-laden flows through confinements accumulate often electrostatic charge, either on purpose or unwanted.
Amid industrial operations, pneumatic conveying electrifies particles the fastest \citep{Kli18}, which frequently leads to hazardous spark discharges.
These discharges continue to be a source of fatal dust explosions around the world, loss of lives, and economic damage.
Only in Germany, about every ten days, one dust explosion occurs caused by static electricity~\citep{Glor03} whereas in Japan 153 industrial accidents were reported over the last fifty years for the same reason \citep{Osh11}.
Electric charge separation in fluidized beds causes operational problems~\citep{Foto17,Foto18} such as the coating of vessel walls~\citep{Sip18}, and is, therefore, an active field of research \citep{Chow19,Tagh19}.
In pharmaceutical devices, eminent electrostatics deteriorates the effectiveness of dry powder inhalers \citep{Wong15}.

On the other hand, the emerging electrostatic forces between particles may modify the flow pattern of the particulate phase \citep{Dho91}.
The effects of this modulation of the particle trajectories are pervasive in technical and environmental flows.
Several industrial applications, such as powder coating, electrostatic precipitation, and electrophotography, function based on this phenomenon.
These applications apply an external electric field to control the particles in the desired way.
In other words, the ability to control the dynamics of electrically charged particle-laden flows brings about the optimization and operational safety of industrial powder processing devices.

Finally, even the climate on earth depends on the electrical forces between sand particles in dust storms.
Thus, more realistic climate models need to account for these forces \citep{Esp16}.

In uncharged powder flows, large turbulent eddies of the carrier gas structure particles in clusters.
As elaborated in the review by Monchaux and Bourgoin \cite{Monch12}, this observation traces back several decades.
Many studies manifested that particles of a Stokes number close to unity reside seldom inside vortices but instead preferably in the saddle regions in-between and at the periphery of the large vortices \citep{Cro88,Kob92,Laz92}.
This phenomenon, termed \textit{turbopheresis}, drives inertial particles toward the wall where turbulent eddies diminish \citep{Cap75,Ree83,Marc02}.

A fluid flows qualitatively different through a rectangular duct compared to a channel.
Channel flows are in a statistical sense one-dimensional, that means all time-averaged properties depend only on the distance to the wall.
In contrast, in duct flows both spatial directions perpendicular to the streamwise coordinate are inhomogeneous and wall-bounded.
These geometrical constraints give rise to secondary flows of Prandtl’s second type which alter the turbulent particle transport compared to a channel \citep{Sha06,Win04}.
These secondary fluid motions occur in turbulent flows through straight ducts of non-circular cross-section and are, therefore, common in engineering applications \citep{Stank16}.
Despite being of relatively low momentum compared to the main flow, the cross-flow secondary motion significantly changes all major parameters of the particle flow, such as concentration, mean velocity, and fluctuations over the cross-section \citep{Noo16}.
Even the locations of the concentration peaks change depending on the precise flow pattern.
In general, due to their distinctive gas flow pattern, powders propagate quantitatively and qualitatively different through rectangular ducts compared to channels.

Since the dynamics of uncharged particles depends on the type of flow, one may suspect a similar influence in case the particles carry an electric charge.
Systematic experimental studies of the effect of a certain charge level seem impossible because one can not impose the same defined charge on all particles in the system.
Therefore, the precise charges of experimentally observed particles remain unknown.
This motivated the usage of numerical techniques, where a clean setting is easily achievable.
However, it is challenging to coupling the underlying mathematical equations, namely the Navier-Stokes equations for the carrier gas flow, Newton's law of motion for the particle dynamics, and Gauss' and Coulomb's law for the electrostatic field.
Therefore, and due to the need for large computational power if important small scale mechanisms shall be resolved, most of the literature on this topic appeared only recently.

One of these recent studies showed through Direct Numerical Simulations (DNS) that for homogeneous isotropic turbulence of a turbulent Reynolds number of about 100, electrostatic forces between particles lead to fluid–particle velocity decorrelation especially in the case of very light particles \citep{Bou20}.
As a consequence, the electrostatic interactions counteract the preferential concentration of particles carrying the same charge and polarity.
The question remains whether this findings are also valid for charged particles in wall-bounded inhomogeneous turbulence that is typically engineering flows. 

For pipe flows at $Re=$~44\,000, one-way coupled Large-Eddy Simulations~(LES) revealed that the electrostatic forces on particles dominate over aerodynamic forces and gravity in the near-wall region \citep{Yao20}.
Also, LES of triboelectric powder charging in pipe flows ($Re=20\,000$--40\,000) \citep{Gro16a,Gro16b,Gro16f,Gro16g} and DNS of the same in channel flows ($Re=3300$) \citep{Gro17a} are available.
But these data are limited to the initial states of charging, thus, the electrostatic charges had a minor impact on the flow patterns. 

On the other hand, LES and DNS showed the strong influence of electrostatic charges on square-shaped duct flows of $Re=$~5000 and 10\,000~\citep{Gro18d,Gro19d,Gro20b}.
While being uncharged, the particles follow the cross-sectional vortical motion of the carrier flow and attain a vortical motion themselves.
When charge was assigned this vortical motion stopped, even though the considered charge was only a fraction of the saturation charge a particle can hold.
The particles still follow the gaseous phase from the center toward the walls, but the electric field pushes them against the inward flux from the wall back to the center of the duct.
As a result, the particle concentration increases in the near-wall region.
The flow pattern of particles carrying a higher charge is unknown due to the lack of simulation data.

All these mentioned simulations enabled a detailed view of the influence of electrostatic charges on powder flows and put forward its importance for the resulting flow pattern.
However, each study focused on specific applications and conditions and their outcomes are, thus, difficult to compare.
For that reason, the relationship between the conveying gas flow pattern on the dynamics of charged particles remains unknown.
In particular, it is unclear whether secondary flows that are induced by the flown-through geometry do affect the distribution of particles.
In this paper, a systematic numerical analysis is presented to elucidate the behavior of identical powders being transported under the same conditions by different flows.
To this end, two flow types are compared which comprise elementary features of complex real pneumatic systems:
A statistically one-dimensional channel flow and a flow through a square-shape duct that forms characteristic vortices in the corners.
The numerical solver is summarized in Section~2, and the detailed simulation conditions and domain geometries are given in Section~3.
Section~4 presents and discusses the results, and Section~5 the concludes the paper.

\section{Mathematical model and numerical methods}

We improved our tool pafiX \citep{Gro20b} to perform the computations presented in this paper.
This section summarizes the mathematical model and numerical methods of pafiX.
The system of governing equations consists of the Navier-Stokes equations describing the motion of the carrier gas and Gauss's law for the electrostatic field which are both given in Eulerian framework.
Further, Newton's law of motion for the particles is solved in Lagrangian manner.
Overall, the continuous and the dispersed phases are four-way coupled.
That means the model accounts for the momentum transfer from the gaseous to the particulate phase and vice versa, and in-between individual particles.

For the constant-density carrier gas flow, the Navier-Stokes equations read
\begin{subequations}
\begin{equation}
\label{eq:mass}
\nabla \cdot {\bm u} \;=\;0\,,
\end{equation}
\begin{equation}
\label{eq:mom}
\frac{\partial {\bm u}}{\partial t} + ({\bm u} \cdot \nabla) {\bm u}
\;=\; - \frac{1}{\rho} \nabla p  + \nu \nabla^2 {\bm u} + {\bm F}_{\mathrm s} + {\bm F}_{\mathrm f} \,.
\end{equation}
\end{subequations}
In these equations, $\bm u$ denotes the fluid velocity, $\rho$ and $\nu$ the gas density and kinematic viscosity, and $p$ the dynamic pressure.
The source term
\begin{equation}
\label{eq:Fs}
{\bm F}_{\mathrm{s}} \;=\; -\dfrac{\rho_{\mathrm{p}}}{\rho} \omega \sum_{i=1}^N {\bm f}_{{\mathrm{fl}},i} 
\end{equation}
accounts for the momentum transfer from the particles to the fluid.
In the above equation, $N$ is the number of particles in a control volume in which the local particle volume fraction is given by $\omega$.
This source term equals the opposite of the sum of the fluid forces, ${\bm f}_{{\mathrm{fl}},i}$, that act on the particles that are located inside the control volume.
The numerical modeling of these fluid forces, which can be separated in drag and lift forces, is described below.

Further, ${\bm F}_{\textrm f}$ represents a constant pressure gradient that balances the fluid's friction loss to the walls.
The component of ${\bm F}_{\textrm f}$ in the downstream direction is $u^2_{\tau} U/A$ and the other components equal zero.
Therein, $u_\mathrm{\tau}$ is the friction velocity and $U/A$ is the ratio between the circumference of the wall wetted by the fluid and cross-sectional area of the flown-through geometry.
Thus, ${\bm F}_{\textrm f}$ directly adjusts the frictional Reynolds number, $Re_\mathrm{\tau} = u_\mathrm{\tau} H / \nu$. 

All spatial derivatives in equations~(\ref{eq:mass}) and~(\ref{eq:mom}) are approximated by second-order accurate central difference schemes.
The temporal derivative in equation~(\ref{eq:mom}) is integrated via an implicit second-order scheme using a variable time-step.

The charge carried by the particles invokes an electric field, ${\bm E}$.
The strength of this field equals the gradient of the electric potential $\varphi_{\rm el}$ which is according to Gauss' law
\begin{equation}
\label{eq:gauss}
\nabla^2 \varphi_{\rm el} \;=\; - \dfrac{\rho_{\mathrm{el}}}{\varepsilon} \, ,
\end{equation}
where ${\rho_{\mathrm{el}}}$ denotes the electric charge density.
Due to the minuscule solid volume fraction in the system, the electric permittivity of the mixture, ${\varepsilon}$, adopts the value of the free space of $\varepsilon=8.85 \times 10^{-12}$ F/m \citep{Rokk10,Rivas07}.
No external electric field is considered, only the field self-induced by charged particles.
Thus, ${\rho_{\mathrm{el}}}$ equals the sum of the charges carried by the particles located within a control volume divided by the size of this volume.
On the other hand, if all particles are uncharged no electric potential arises and equation~(\ref{eq:gauss}) is not solved.
Moreover, since the electric potential at the walls equals zero (see next section) the above equation can be solved using periodic boundary conditions in the homogeneous flow directions.
The left-hand side of Gauss's law \eqref{eq:gauss} is discretized using second-order central difference schemes.

The acceleration of each particle is solved separately according to Newton's second law of motion, i.e.,
\begin{equation}
\label{eq:newton}
m_\mathrm{p} \, \dfrac{\mathrm{d} \bm u_{\textrm p}}{\mathrm{d} t} \;=\; {\bm f}_{\mathrm{d}} + {\bm f}_{\mathrm{l}} + {\bm f}_{\mathrm{coll}} + {\bm f}_{\mathrm{el}} + {\bm f}_{\mathrm{g}}\, .
\end{equation}
Therein, ${\bm f}_{\mathrm{d}}$ and ${\bm f}_{\mathrm{l}}$ are the drag and lift forces on the particle due to the surrounding gas.
The sum of these two forces is ${\bm f}_{\mathrm{fl}}$ in equation~(\ref{eq:Fs}).
Further, ${\bm f}_{\mathrm{coll}}$ is the force due to particle-particle or wall-particle collisions, ${\bm f}_{\mathrm{el}}$ due to electrostatic forces, and ${\bm f}_{\mathrm{g}}$ due to gravitation.

The acceleration due to the drag force is computed according to the law by Putnam \cite{Put61}.
The lift force is modelled following to the analysis of Saffmann \cite{Saf65,Saf68} and the correction of Mei \cite{Mei92} for particle Reynolds numbers higher than the shear Reynolds number.
A variant of the hard-sphere approach, namely the ray casting method~\citep{Roth82,Schr01} which we adapted to handle spherical particles, accounts for collisional forces between particles.
During wall collisions, the particle's wall-normal velocity component changes its sign and the tangential velocity component of the particle remains constant.

Finally, the electrostatic force on a particle carrying the charge $Q$ is computed by the hybrid approach \citep{Gro17e}.
The hybrid approach superimposes the Coulombic interactions between the particle and its $N$ neighboring particles, ${\bm f}_{\mathrm{el,C}}$, to the far-field forces calculated with Gauss law, ${\bm f}_{\mathrm{el,G}}$,
\begin{equation}
\label{eq:fel}
{\bm f}_\mathrm{el} = {\bm f}_\mathrm{el,C} + {\bm f}_\mathrm{el,G}
\quad
\text{with}
\quad
{\bm f}_\mathrm{el,C} = \sum\limits_{n=1}^N \dfrac{Q \, Q_n \, {\bm z}_n}{4 \, \pi \, \varepsilon \, |{\bm z}_n|^{3}} 
\quad
\text{and}
\quad
{\bm f}_\mathrm{el,G} = - Q \, \nabla \varphi_{\rm el} \, . 
\end{equation}
The charge of the neighboring particles is $Q_n$ and the vector ${\bm z}_n$ points from their center point to the center of the particle under consideration.
As stated above, the electric potential, $\varphi_{\rm el}$, arises only from the charge carried by the particles since we consider no external field.

At each numerical time-step ($t_1$), equation~(\ref{eq:newton}) is solved by the Euler-forward method.
This method is explicit, thus, the fluid flow and the electric field are required at the previous time-level $t_0$.
Then, the trajectories of the particles are integrated from $t_0$ to $t_1$ through the implicit second-order accurate Crank-Nicolson method.
Afterward, the fluid flow field at $t_1$ is computed in an implicit manner as described above.
During each iteration of equation~(\ref{eq:mom}), the momentum source term ${\bm F}_s$ on the right hand side is updated to reflect the change of the flow field.
Thus, the gaseous and the particulate phase are implicitly coupled to each other.

\section{Numerical set-up and simulation conditions}

\begin{figure*}
\centering
\subfigure[]{
\begin{minipage}[t]{0.2\textwidth}
\vspace{-55mm}
\renewcommand{\arraystretch}{0.8}
\begin{tabular}{ccc}
 BC & channel & duct \\
\hline 
 $x$ & periodic & periodic \\ 
 $y$ & wall & wall \\ 
 $z$ & periodic & wall\\ 
\end{tabular}
\renewcommand{\arraystretch}{1.0}
\end{minipage}
\hspace{-30mm}
\begin{minipage}[t]{0.2\textwidth}
\begin{tikzpicture}[scale=.65]
 \draw [->, thick] (0,-.75) -- (1,.25) node [below,xshift=5] {$x$};
 \draw [->, thick] (0,-.75) -- (0,.25) node [below,xshift=5] {$y$};
 \draw [->, thick] (0,-.75) -- (1,.-.75) node [below,xshift=5] {$z$};
\end{tikzpicture}
\end{minipage}
\hspace{-10mm}
\begin{tikzpicture}[scale=.65]
 \draw [thick,fill=gray,fill opacity=.8] (0,-1.5) -- (4,-1.5) -- (10,4.5) -- (6,4.5) -- (0,-1.5);
 \draw [thick,fill=gray,fill opacity=.8] (0,0) -- (4,0) -- (10,6) -- (6,6) -- (0,0);
 \draw [thick,|-|] (-.5,-1.5) -- node [above,rotate=90] {$H$} (-.5,0);
 \draw [thick,|-|] (-.3,.2) -- node [above,rotate=45] {$6\,H$} (5.7,6.2);
 \draw [thick,|-|] (0,-1.9) -- node [below,rotate=0] {$2\,H$} (4,-1.9);
\end{tikzpicture}
\label{fig:domainChannel}}
\hspace{-25mm}
\subfigure[]{
\begin{tikzpicture}[scale=.65]
 \draw [thick,fill=gray,fill opacity=.8] (0,-1.5) -- (6,4.5) -- (6,6)-- (0,0) -- (0,-1.5);
 \draw [thick,fill=gray,fill opacity=.8] (0,-1.5) -- (1.5,-1.5) -- (7.5,4.5) -- (6,4.5) -- (0,-1.5);
 \draw [thick,fill=gray,fill opacity=.8] (0,0) -- (1.5,0) -- (7.5,6) -- (6,6) -- (0,0);
 \draw [thick,fill=gray,fill opacity=.8] (1.5,-1.5) -- (7.5,4.5) -- (7.5,6)-- (1.5,0) -- (1.5,-1.5);
 \draw [thick,|-|] (-.5,-1.5) -- node [above,rotate=90] {$H$} (-.5,0);
 \draw [thick,|-|] (-.3,.2) -- node [above,rotate=45] {$6\,H$} (5.7,6.2);
 \draw [thick,|-|] (0,-1.9) -- node [below,rotate=0] {$H$} (1.5,-1.9);
\end{tikzpicture}
\label{fig:domainDuct}}
\caption{Dimensions ($H=4$~cm) and boundary conditions of the (a) channel and (b) duct flow domains.}
\label{fig:domain}
\end{figure*}
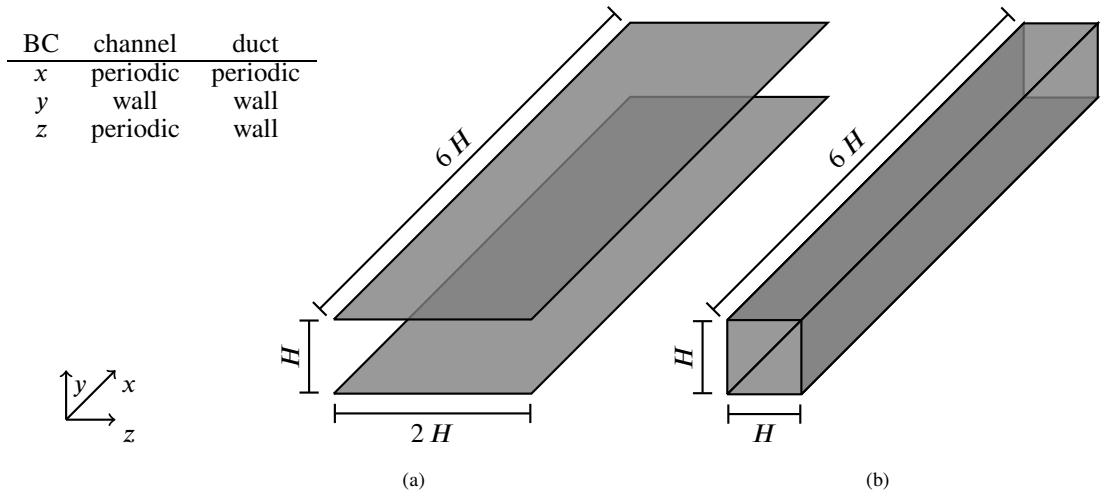

We simulated particle-laden flows through two distinctive geometries:
a channel and square-shaped duct, as sketched in figure~\ref{fig:domain}.
The pressure gradient imposed to drive the flow of the carrier gas is chosen so the frictional Reynolds number of all cases equals 360.
These geometries are both fundamental and generic, thus, representing elements of numerous complex practical engineering applications.
The flow structures in both geometries are qualitatively different:
a channel is an ideal model flow for the analysis of wall-bounded turbulence whereas a duct flow intrigues due to its transverse (secondary) mean motion.
Therefore, a comparison of particle-laden flows through both domains allows us to draw concrete conclusions regarding the influence of the carrier flow pattern on the effect of electrostatic forces.

The complete simulation set-up, i.e., conditions, mesh, and dimensions, aims to be as consistent as possible between both flow types.
Doing so ensures that physics and not numerics causes any observed difference in the results.
The powder flows vertically downward, in other words, both the gravity vector and the constant pressure gradient in equation~(\ref{eq:mom}) point in the positive $x$-direction.
For the channel, periodic boundary conditions mimic the two homogeneous directions, i.e.~$x$ and $z$, and solid walls confine the flow in the $y$-direction.
The dimensions of the channel equal $6\,H \times H\,\times\,2\,H$ ($H=4$~cm) are sufficiently large to capture the main flow structures for the herein considered Reynolds number \mbox{\citep{Mos99}.
Thus, the simulated fluid flow is independent of the domain size.
For particles of high Stokes numbers, which are mostly driven by low frequency fluid motions, these dimensions might be too small to obtain domain independent results.
For example, the particle concentrations alters locally up to 20\% if the domain size is increased by a factor of three in $x$ and $z$-direction \mbox{\citep{Sar12}}.}
However, since we consider low Stokes numbers, and for the sake of computational efficiency, we decided for the above given dimensions.

The set-up of the duct is identical to the channel with exception of the $z$-direction where the domain size is $H$, so the duct is of a square cross-section, and wall boundary condition is applied.  
At each wall, no-slip is assumed for the velocity and zero-Dirichlet boundary condition is prescribed for the electric potential $\varphi_{\rm el}$, i.e., the walls are assumed to be grounded and fully conductive.

Also, the meshes for both geometries are chosen as consistent as possible to optimize the comparability of the simulations.
The points are distributed uniformly in all homogeneous directions.
In the inhomogeneous directions (i.e., $y$ for the channel and $y$ and $z$ for the duct) the grids are refined toward the walls.
The refinement is accomplished by a cosine stretching function \citep{Kim87} where the $N$ grid points in $y$-direction are located at $y_j/H=\cos \left(\pi (j-1) / (N-1)\right)/2$ with $j=1,\ldots,N$.
For the duct, the analogous equation defines the location of the grid points in the $z$-direction.

In Eulerian-Lagrangian simulations, the grid is a compromise between the wish to resolve the gas phase as good as possible and the underlying assumption of each cell being larger than a particle.
All cases are computed on $256 \times 144 \times 144 = 5\,308\,416$ grid points.
Compared to classical DNS, this grid consists of 2.5 times more points than the one of Kim et al. \cite{Kim87} for channel flows for the same Reynolds number, and 5.4~times more cells than the one of Huser and Biringen \cite{Hus93b} for duct flows of a 66\% higher Reynolds number.
The difference in the mean and rms flow quantities of channel flow of $Re_\tau =360$ is less than 1\% when computed on grids consisting of $256 \times 128 \times 128$ and $512 \times 288 \times 288$ grid points \citep{Vre14}.
Therefore, and due to the fact that inertial particles, which are the interest of the present study, act as low pass filters and are not affected by very small scale fluctuations, the grid of $256 \times 144 \times 144$ points is suitable for the present investigation.
This results in a uniform mesh spacing of 8.4~wall units in the streamwise direction and of 5.0~wall units in the spanwise direction of the channel flow.
In the inhomogeneous directions, the first grid point is located at a distance of 0.0044~wall units from the boundary.

\setlength{\tabcolsep}{8pt}
\renewcommand{\arraystretch}{1.2}
\begin{table}[tb]
\centering
\caption{Parameters of the powder flows investigated through the simulations.
Each case is run in a channel and in a duct, once with uncharged ($S\!t_\mathrm{el}=0$) and once with charged particles.}
\label{tab:param}
\medskip
\begin{tabular}{cccccc}
\hline
\hline
Case & $\rho_N^+$ & $V_\mathrm{p}/V$ & $\rho_\mathrm{p}/\rho$ & $S\!t$ & $S\!t_\mathrm{el}$ \\
\hline
1 & $1.0\times10^8$~m$^{-3}$ & $1.77\times10^{-7}$ &    7900 &  8.0 & 0.002  \\
2 & $1.0\times10^8$~m$^{-3}$ & $1.77\times10^{-7}$ & 31\,600 & 32.0 & 0.004  \\
3 & $2.5\times10^7$~m$^{-3}$ & $4.42\times10^{-8}$ &    7900 &  8.0 & 0.001  \\
4 & $4.0\times10^8$~m$^{-3}$ & $7.07\times10^{-7}$ &    7900 &  8.0 & 0.004  \\
\hline
\hline
\end{tabular}
\end{table}
\setlength{\tabcolsep}{1pt}
\renewcommand{\arraystretch}{1.0}

To avoid size-distribution effects, the present study considers a monodisperse powder where each particle has a diameter of 15~microns, respectively 0.135 wall units.
That means, the volume of the smallest cell of the Eulerian grid is 144~times larger than the volume of a particle.
Table~\ref{tab:param} summarizes the conditions of the four investigated cases.
Further, in Table~\ref{tab:param}, $\rho_N$ denotes the particle number density, $V_\mathrm{p}/V$ the related dimensionless particle/gas volume ratio, and $\rho_\mathrm{p}/\rho$ the particle/gas density ratio.
The absolute number of particles in the system, which can be derived from $\rho_N$ and the domain size, varies between 9600 for the duct flow of case~3 and 307\,200 for the channel flow of case~4.
The particle dynamics is characterized by two different Stokes numbers.
The first one is defined as the ratio of the particle response time-scale to the flow time-scale, i.e.,
\begin{equation}
S\!t=\tau_\mathrm{p}/\tau_\mathrm{f} \, .
\end{equation}
The second one is the electrical Stokes number, given by the ratio of the particle response time-scale to a time-scale characterizing the Coulombic particle interaction \citep{Bou20},
\begin{equation}
S\!t_\mathrm{el}=\tau_\mathrm{p}/\tau_\mathrm{el} \, .
\end{equation}
The values of $S\!t_\mathrm{el}$ in Table~\ref{tab:param} are based on time-scales defined as
\begin{align}
\tau_\mathrm{p} \propto 2 \rho_\mathrm{p} r^2 / (9 \rho \nu)\, , \qquad
\tau_\mathrm{f} \propto \nu / u^2_\mathrm{\tau} \, , 
\qquad \text{and} \qquad
\tau_\mathrm{el} \propto \dfrac{2}{Q} \left(\dfrac{\pi\,\varepsilon\,m_\mathrm{p}}{\rho_N}\right)^{1/2} \, .
\end{align}
The time-scale $\tau_\mathrm{p}$ quantifies the response of a spherical particle to a Stokes flow \citep{Som12}, $\tau_\mathrm{f}$ the gas flow's frictional motion, and $\tau_\mathrm{el}$ the acceleration of a charged particle by the electric field surrounding an equally charged particle.
The definition of $\tau_\mathrm{el}$ involves the assumption that the other particle is located at a distance which equals the spacing between particles if they would be uniformly distributed in the domain.

According to Table~\ref{tab:param}, the effect of electrical forces is analyzed for four different cases:
In cases~1 and~2 the particle number density, respectively the particle/gas volume fraction, is the same whereas the particle/gas material density ratio is increased by a factor of~4.
On the other hand, in cases~1, 3, and~4 the particle/gas material density ratio is the same, but the particle number density is in case~3 decreased by a factor of~4 and in case~4 increased by a factor of~4.
Each of these four cases is simulated once for a channel and once for a duct flow.

Additionally, each case is once run with uncharged particles and once where a charge of 1.26~fC is assigned to each particle.
This value represents approximately 1\% -- 2\% of the maximum charge, also called equilibrium or saturation charge, a particle of the considered size can hold following the summary of Matsuyama \cite{Mat18}.
The charge was chosen that low to investigate its influence on the powder flow.
By doing so they remain airborne.
For higher charges, they tend to stick or bounce repeatedly on the walls.
This amount of charge is assumed to be constant during the simulation and does not change even during particle-particle or wall-particle collisions.
In other words, we compare duct to channel flows and uncharged to weakly charged flows.
Regarding modeling of the build-up of particle charge, the reader is referred to the review of Chowdhury et al.~\mbox{\cite{Chow21}}.

\section{Resulting particle concentration and dynamics}

\begin{figure*}[tb]
\centering
\subfigure[]{\includegraphics[trim=0mm 0mm 0mm 0mm,clip=true,width=0.52\textwidth]{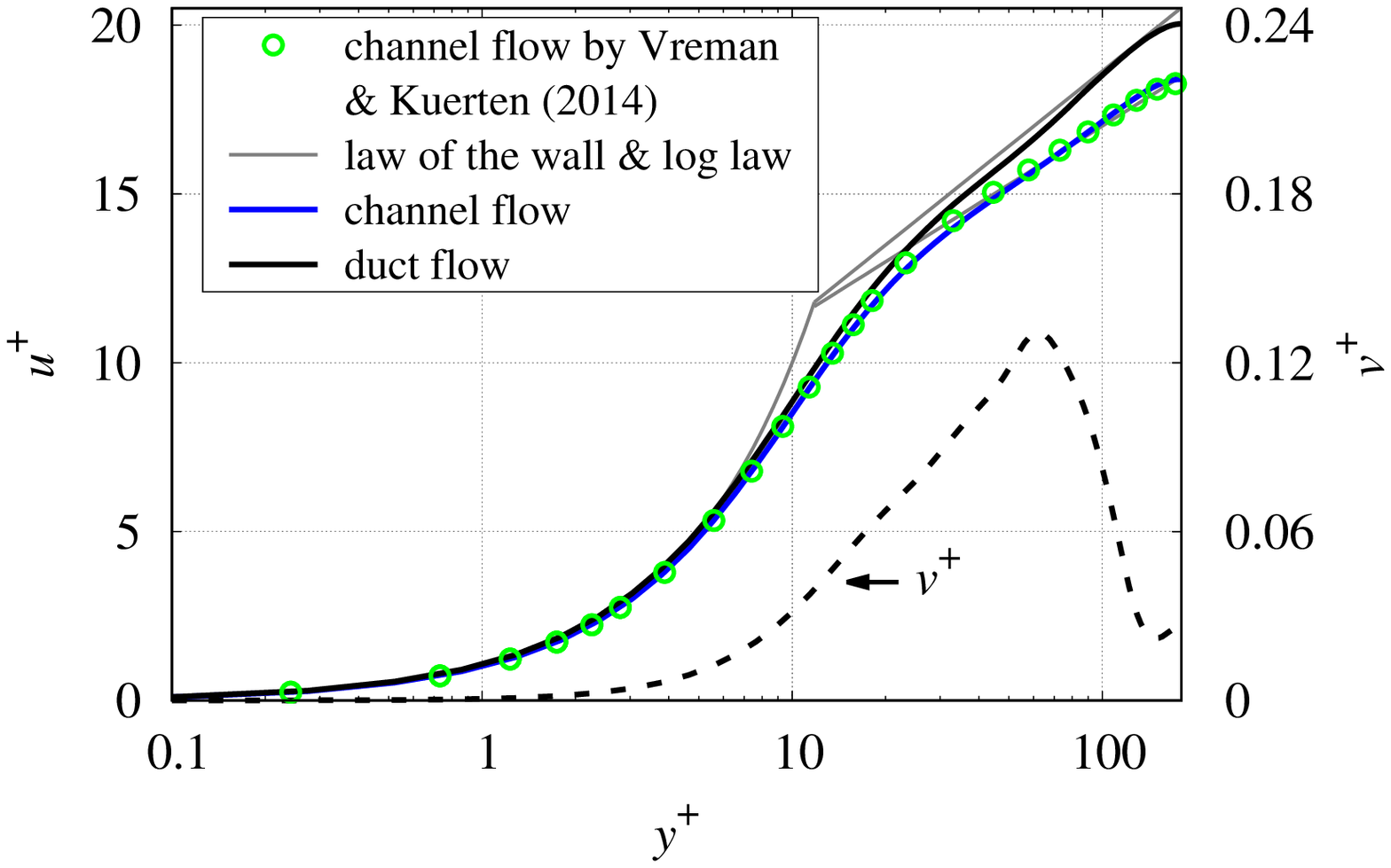}\label{fig:ugas}}
\subfigure[]{\includegraphics[trim=0mm 0mm 0mm 0mm,clip=true,width=0.48\textwidth]{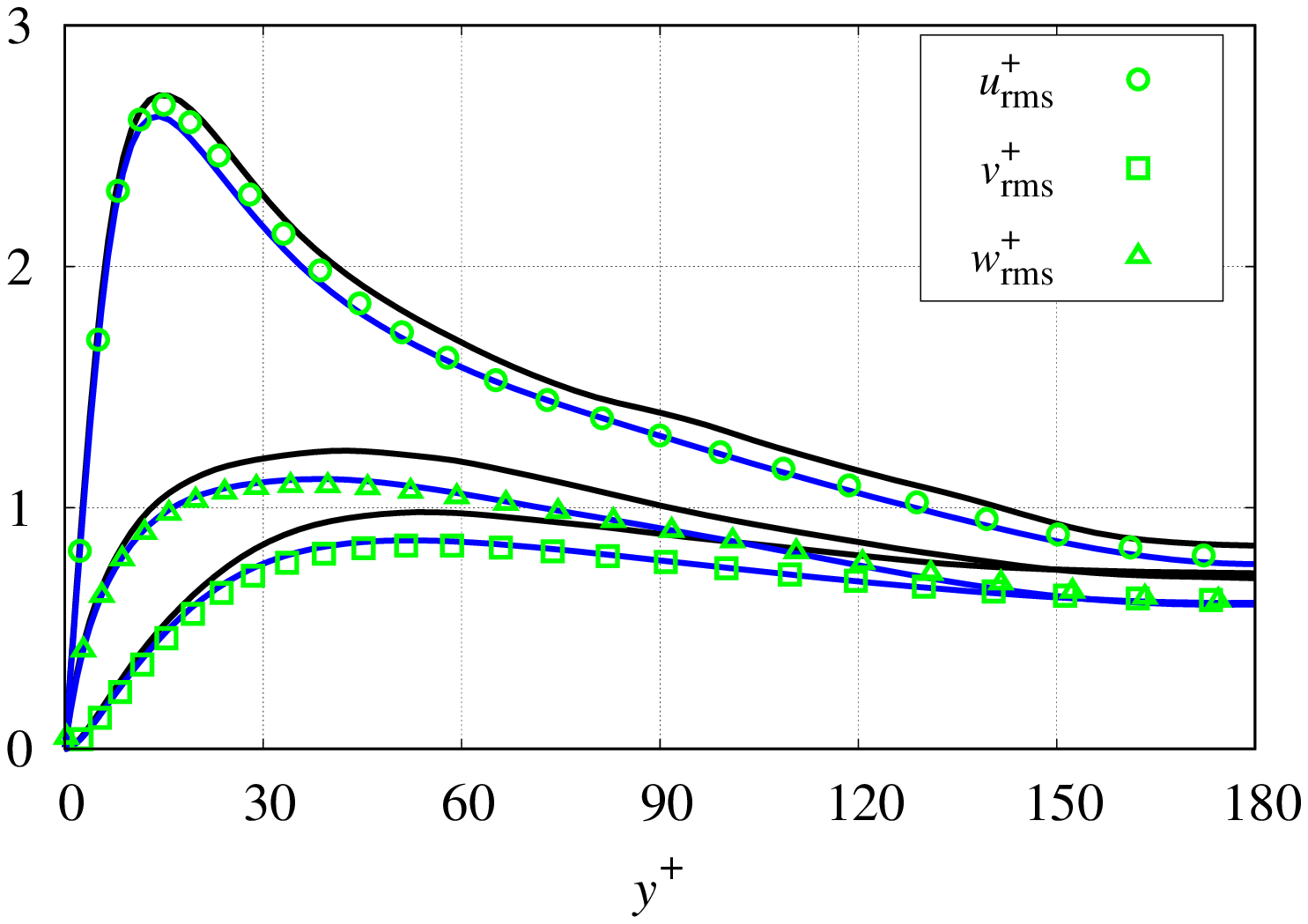}\label{fig:rms}}\\
\caption{Comparison of (a) the time-averaged velocities and (b) the fluctuations of the carrier gas flow without particles of the present simulations to highly-resolved channel flow DNS from the literature and the law of the wall and log law.
The log law is presented in the form $u^+=2.5 \ln y^+ + 5.5$ for a channel flow and in the form $u^+=3.2 \ln y^+ + 3.9$ for a square duct flow \citep{Sha06}.
In (a), additionally the time-averaged wall-normal velocity of the duct flow due to secondary flows is given in the dashed line.
The spatial axis ($y^+$) of the duct flow data connects the bisector of a wall with the centreline of the duct.
All profiles are normalized to the mean friction velocity.
}
\label{fig:ugasrms}
\end{figure*}

The simulation of the carrier gas flow is validated with DNS data from the literature, see figure~\ref{fig:ugasrms}.
Figure~\ref{fig:ugas} gives the time-averaged streamwise velocity profiles, and figure~\ref{fig:rms} the root-mean-square of the three components of the velocity fluctuations.
The data provided therein is non-dimensionalized with the frictional quantities, as indicated by the superscript '+'.
More specifically, the spatial coordinate ($y^+$) is given in wall units, and the velocity components are normalized with $u_\tau$.
In these plots and the remainder of this paper, the blue curves and symbols represent the channel flow and the black ones the duct flow.
Both the average velocity and the fluctuations of our channel flow simulations match the DNS for the same conditions of Vremen and Kuerten \cite{Vre14}, which were computed, as mentioned above on $512 \times 288 \times 288$ grid points.
Also, our predicted average streamwise velocity profile matches the law of the wall and the log law.

The duct flow simulations were already validated by Grosshans et al.~\cite{Gro20b}.
But compared to our previous simulations, the present calculations concern a lower Reynolds number and are run on an even finer grid.
For historical reasons, duct flows are often simulated for $Re_\tau = 600$ or a multiple of that and then compared to the pioneering work of Huser and Biringen \cite{Hus93b} or later works.
However, for answering the research question raised in the present paper, it is the priority to consider for the channel and duct flow the same Reynolds number, i.e., $Re_\tau =360$.
Even more important, the highest priority of our set-up is to optimize the comparability between the duct and channel flow simulations, which is a strong argument for the usage of a consistent mesh.
The profiles for the duct are plotted along the axis from the bisector of one wall to the centreline of the duct.
The average streamwise velocity profile for the duct matches the modified law of the wall and the log law.

The presence of a net fluid flux due to secondary flow from the wall toward the center of the duct is the fundamental difference from the channel flow where this velocity component equals zero on average.
In figure~\ref{fig:ugas}, besides the streamwise velocities, the time-averaged wall-normal velocity of the duct flow is depicted.
The magnitude of the secondary flow is approximately two magnitudes lower than that of the primary flow, as indicated by the different scaling of the velocity axes.
The question is whether such a small flow alters the transport of particles significantly.

As regards the particulate phase, first the convergence behavior is analyzed.
To this end, the particles are initially seeded at random positions in the fully developed carrier gas flow.
At seeding, the initial particle velocity equals the velocity of the surrounding fluid.
After their release, the particles start to migrate until at some point their time-averaged concentration over the channel's and duct's cross-section converges.
In other words, during the instationary part of the simulation, the powder distribution develops from initially homogeneous to statistically stationary.

\begin{figure*}[tb]
\centering
\subfigure[Case 1: $S\!t=8$, $S\!t_\mathrm{el}=0.002$] {\includegraphics[trim=0mm 6mm 0mm 0mm,clip=true,width=0.47\textwidth]{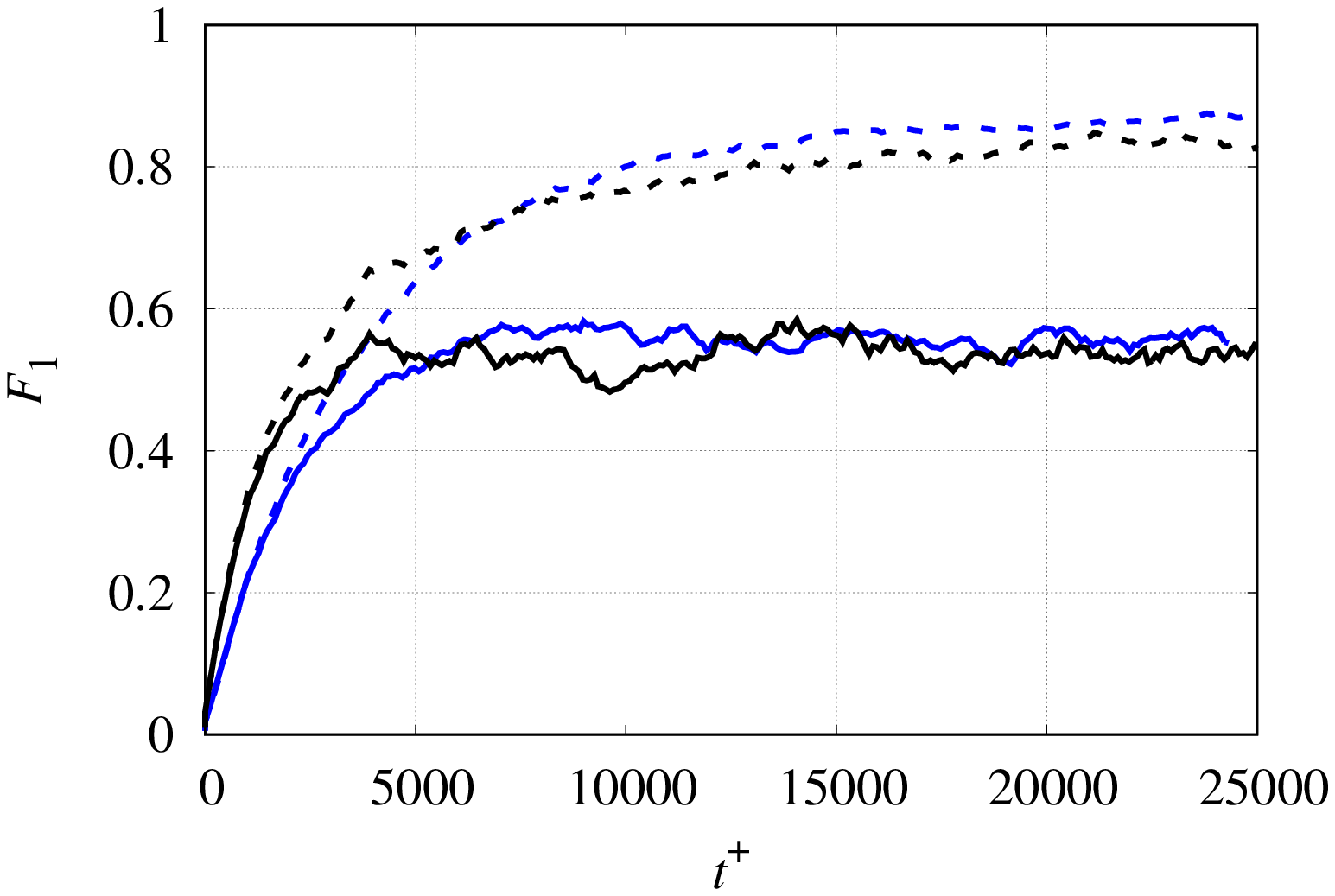}\label{fig:convergence1}}\qquad
\subfigure[Case 2: $S\!t=32$, $S\!t_\mathrm{el}=0.004$]{\includegraphics[trim=0mm 6mm 0mm 0mm,clip=true,width=0.47\textwidth]{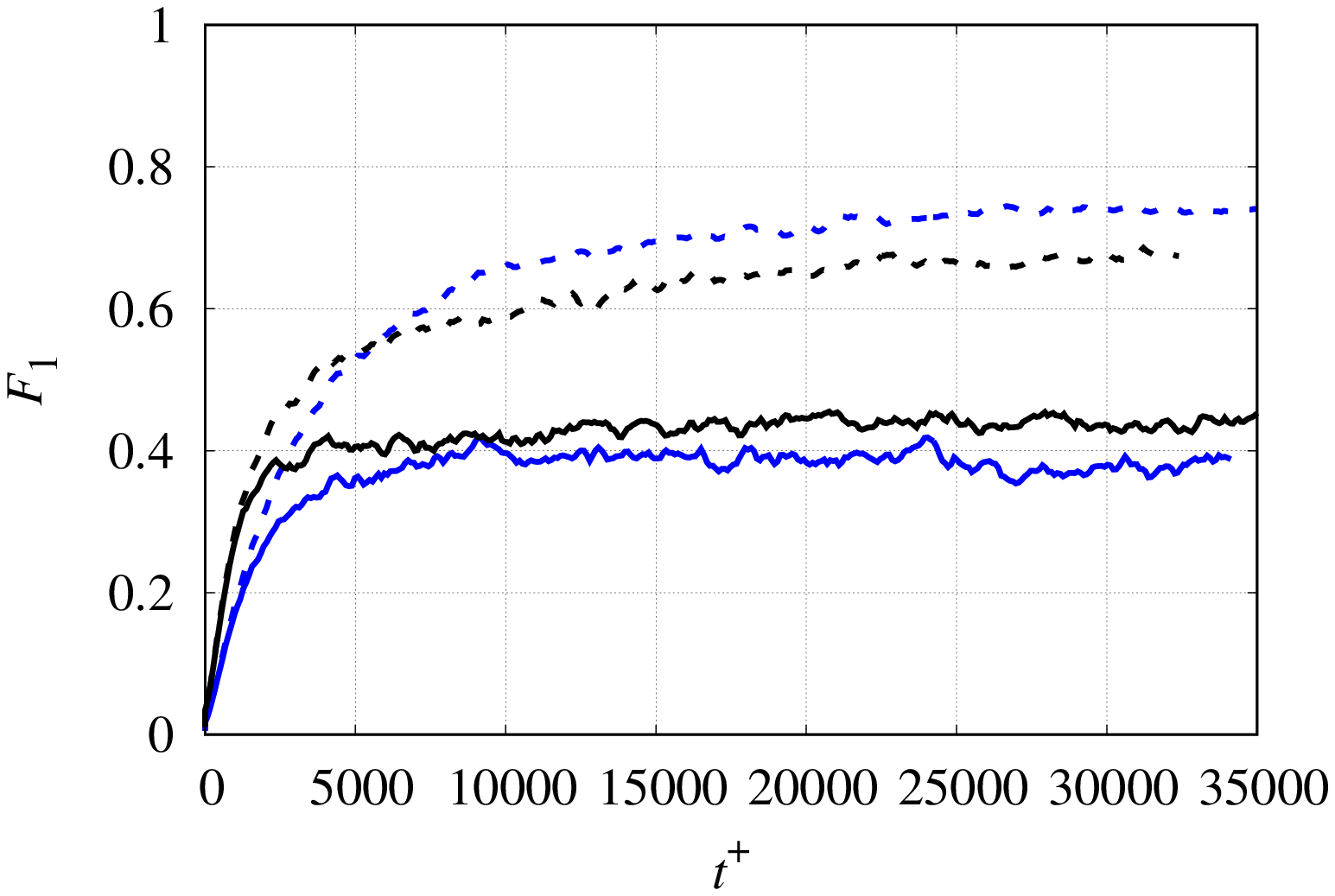}\label{fig:convergence2}}\\
\subfigure[Case 3: $S\!t=8$, $S\!t_\mathrm{el}=0.001$] {\includegraphics[trim=0mm 6mm 0mm 0mm,clip=true,width=0.47\textwidth]{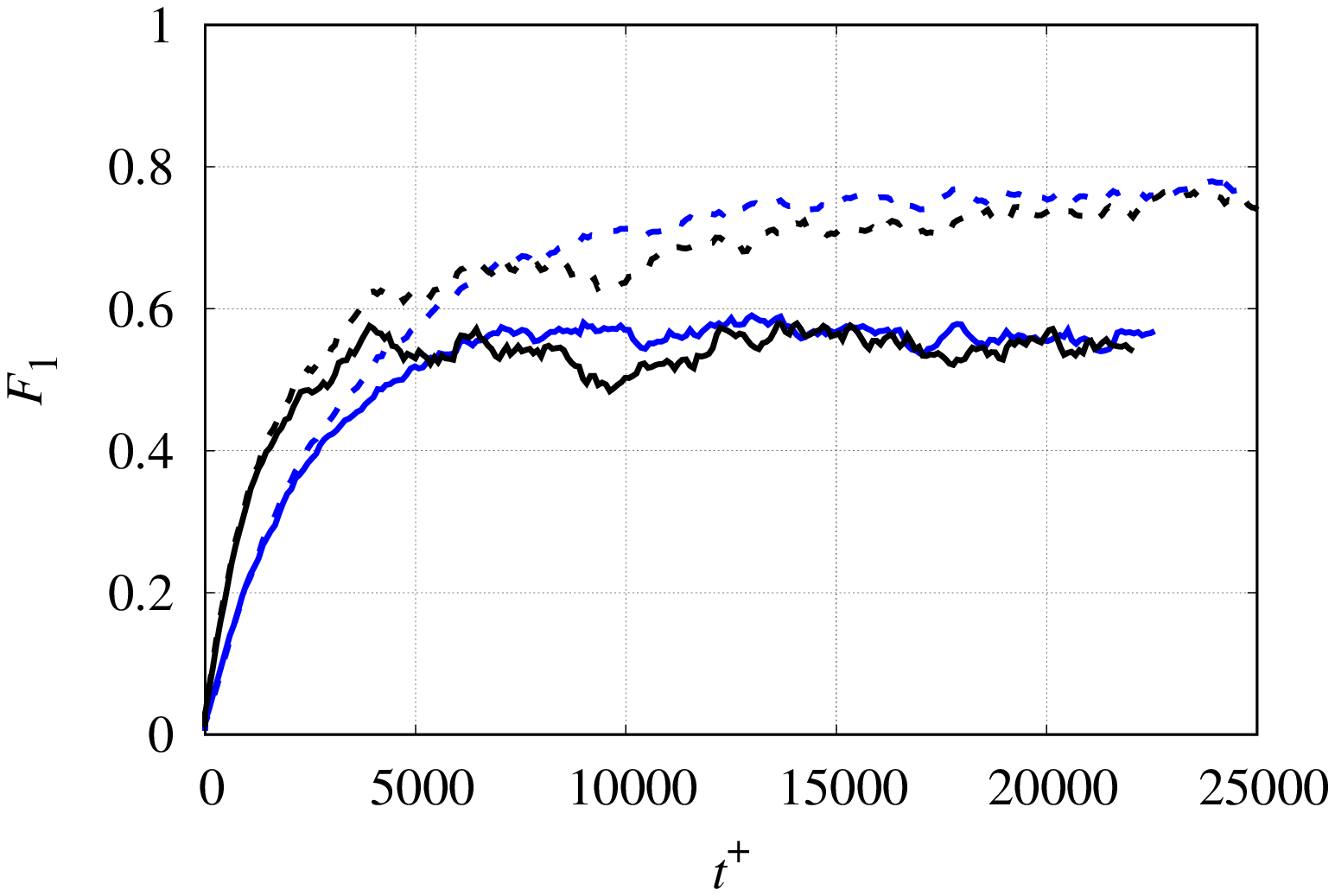}\label{fig:convergence3}}\qquad
\subfigure[Case 4: $S\!t=8$, $S\!t_\mathrm{el}=0.004$] {\includegraphics[trim=0mm 6mm 0mm 0mm,clip=true,width=0.47\textwidth]{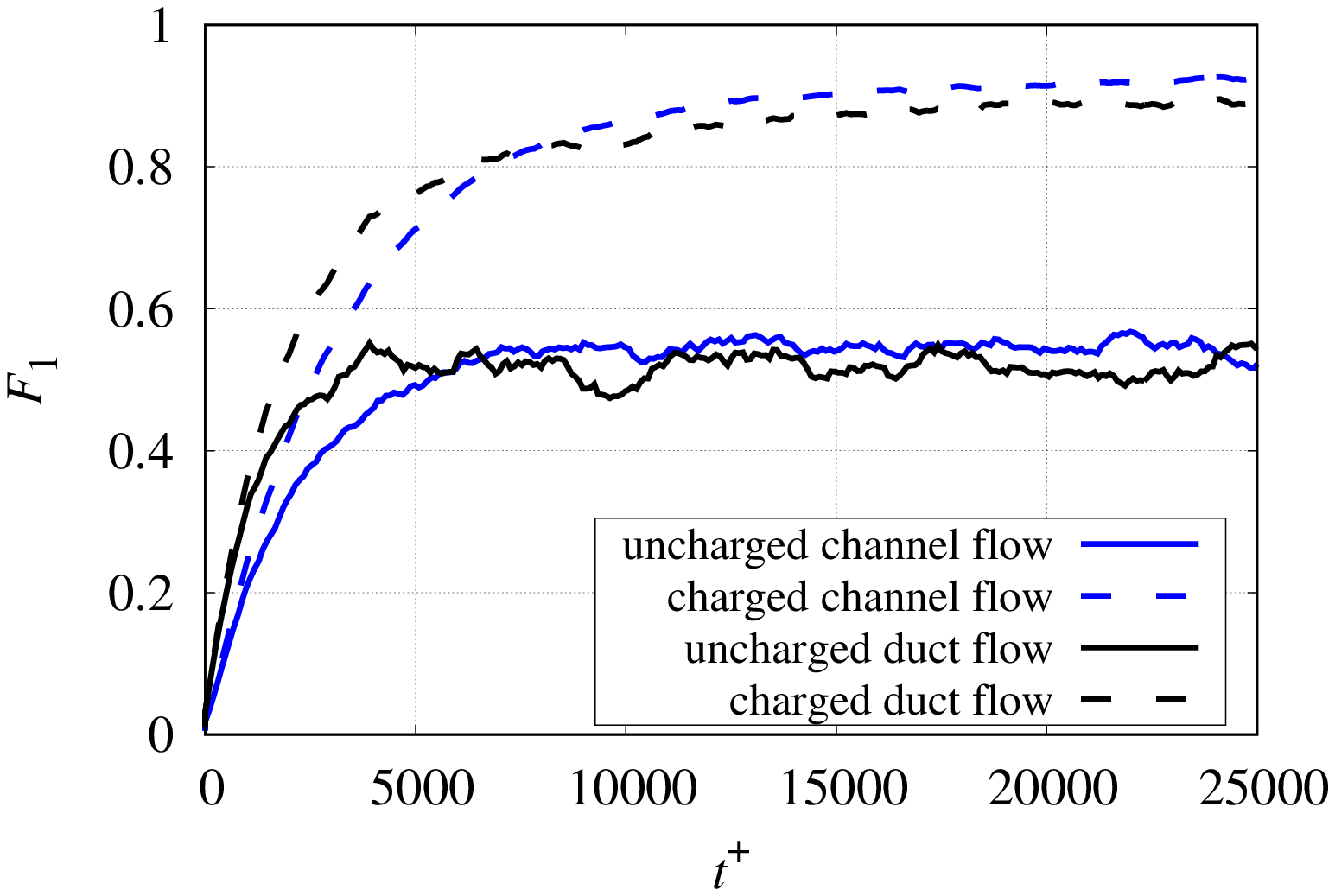}\label{fig:convergence4}}
\caption{Temporal evolution of the fraction of particles located less than one viscous length-scale from a wall.
}
\label{fig:convergence}
\end{figure*}

Figure~\ref{fig:convergence} plots this instationary evolution for all cases.
The blue curves and symbols represent the channel flow and the black curves the duct flow.
Additionally, the uncharged cases are plotted in solid and the charged cases with dashed lines.
The temporal axis is non-dimensionalized with frictional quantities, $t^+=t \, u^2_\tau / \nu$.
The fraction $F_1$ denotes the number of particles residing within one frictional lengthscale from a wall ($y^+ \le 1$) divided by the total number of particles.
These particles are captured in the viscous sublayer.

\begin{figure*}[tb]
\centering
\subfigure[Case 1: $S\!t=8$, $S\!t_\mathrm{el}=0.002$] {\includegraphics[trim=0mm 6mm 0mm 0mm,clip=true,width=0.47\textwidth]{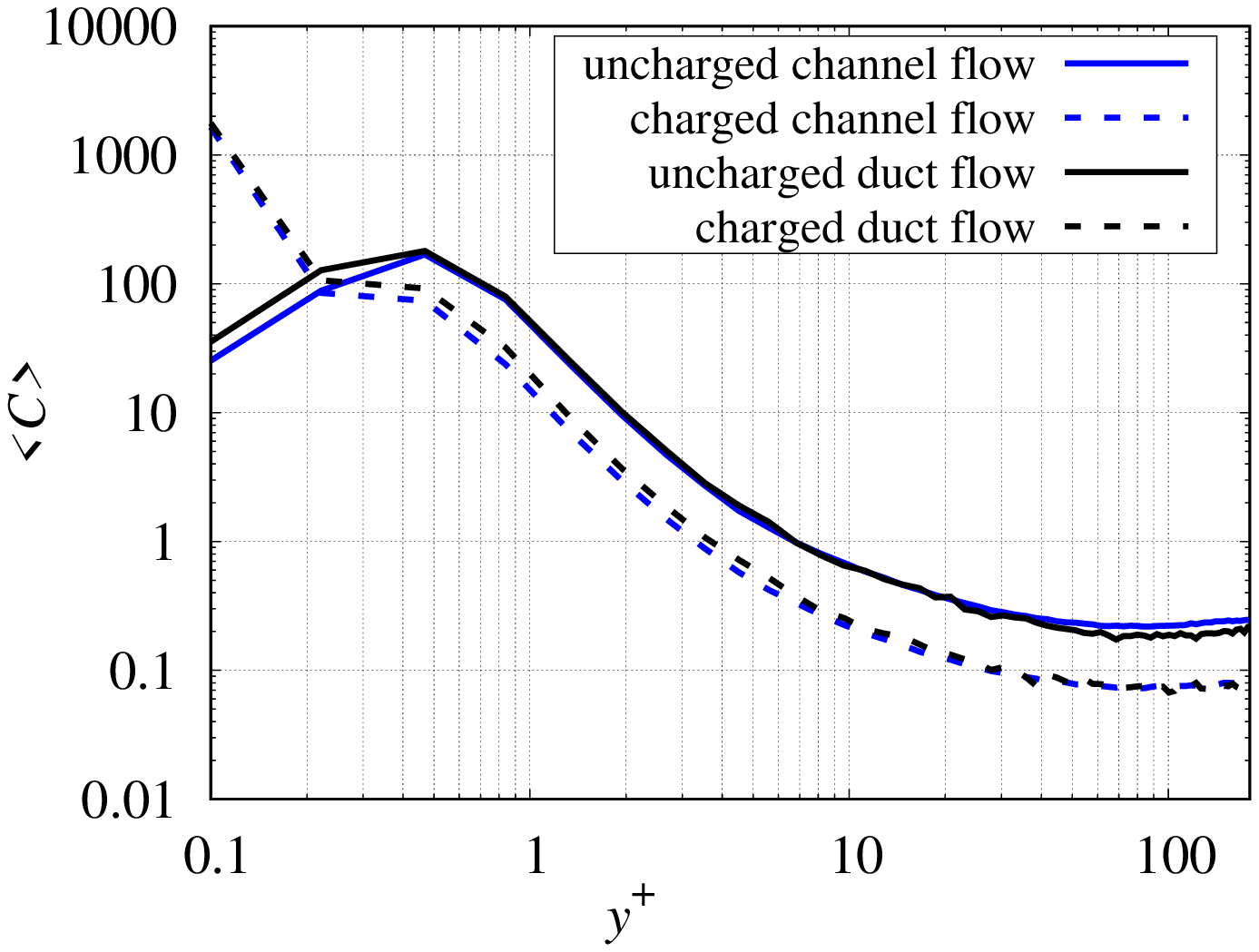}\label{fig:np1}}\qquad
\subfigure[Case 2: $S\!t=32$, $S\!t_\mathrm{el}=0.004$]{\includegraphics[trim=0mm 6mm 0mm 0mm,clip=true,width=0.47\textwidth]{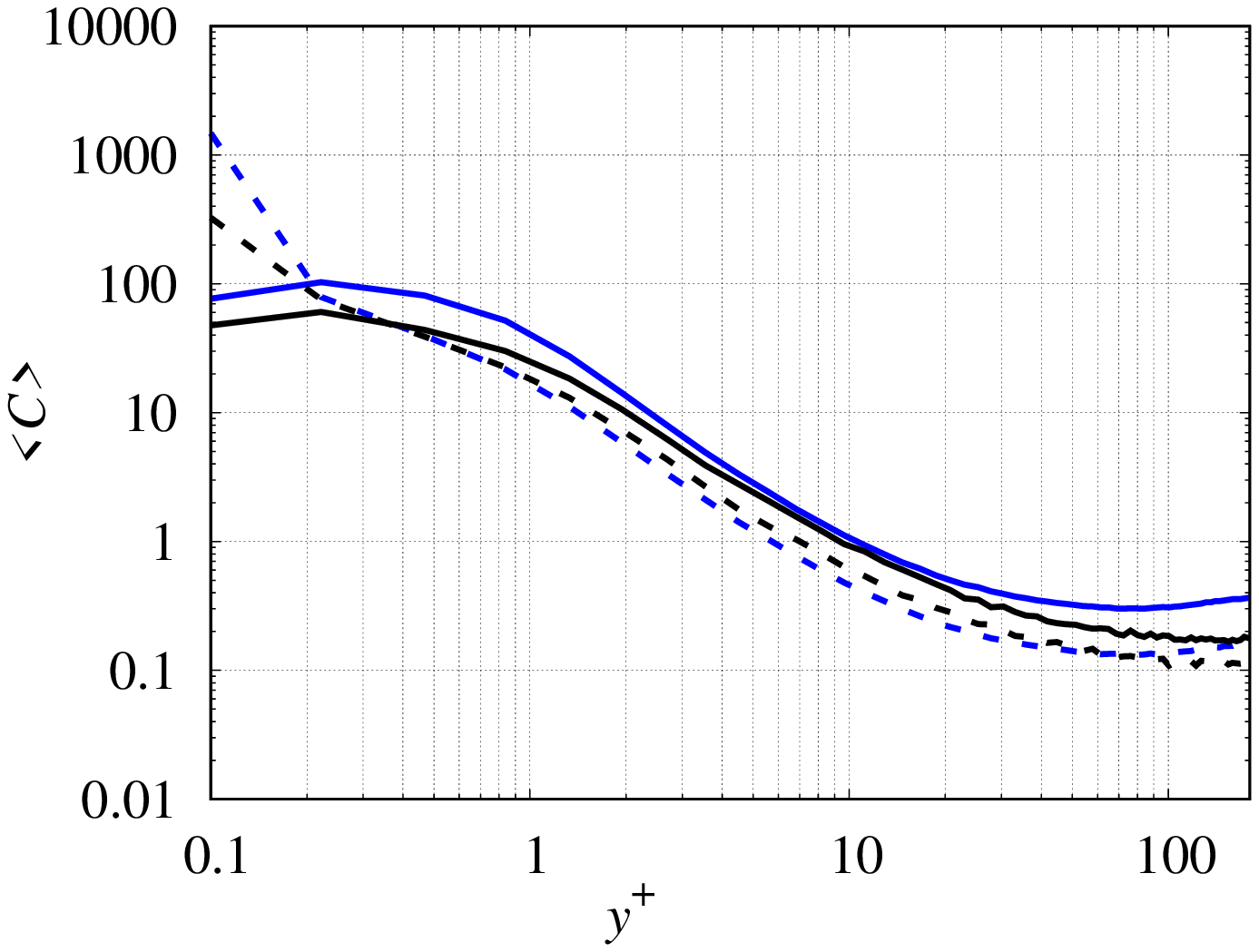}\label{fig:np2}}\\
\subfigure[Case 3: $S\!t=8$, $S\!t_\mathrm{el}=0.001$] {\includegraphics[trim=0mm 6mm 0mm 0mm,clip=true,width=0.47\textwidth]{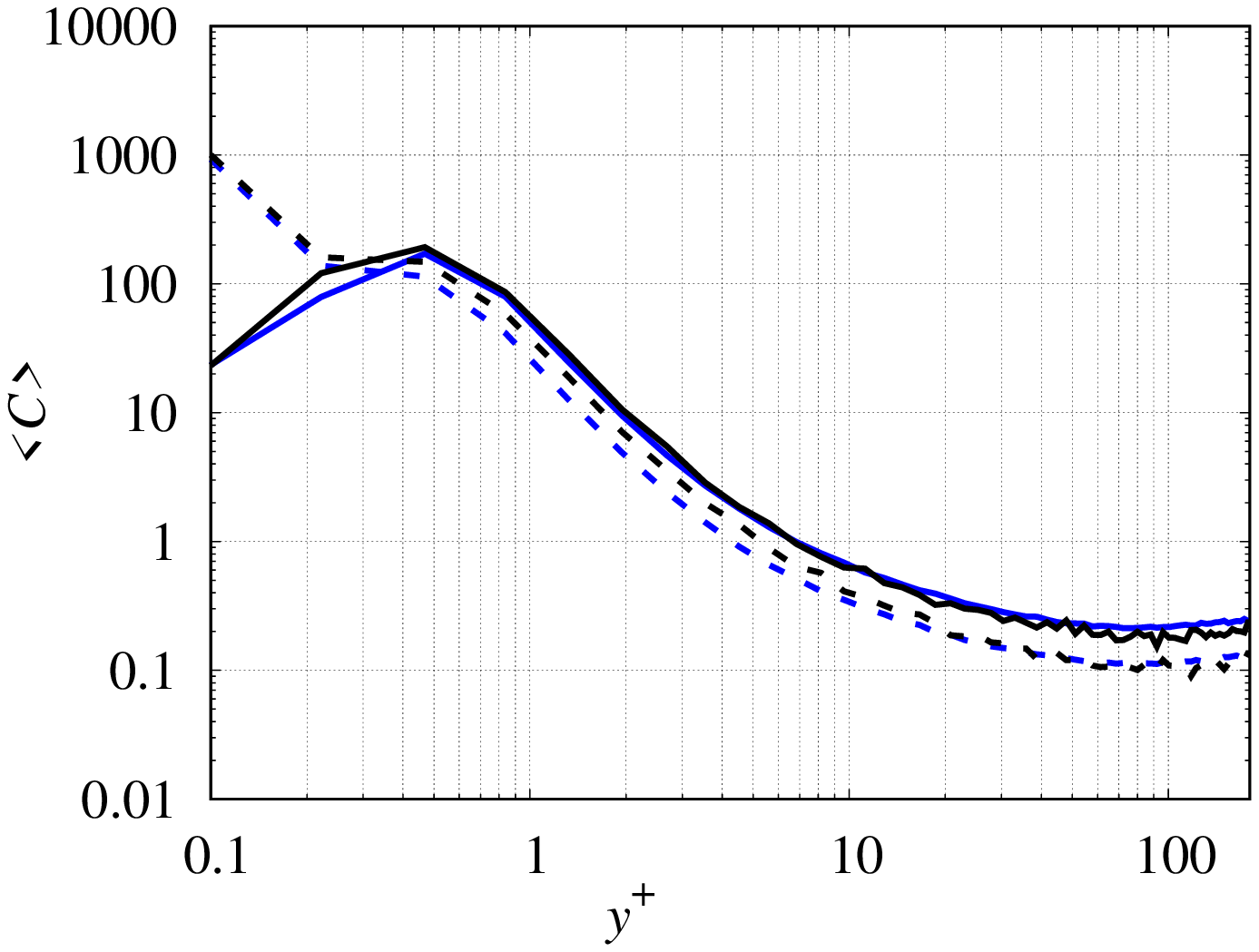}\label{fig:np3}}\qquad
\subfigure[Case 4: $S\!t=8$, $S\!t_\mathrm{el}=0.004$] {\includegraphics[trim=0mm 6mm 0mm 0mm,clip=true,width=0.47\textwidth]{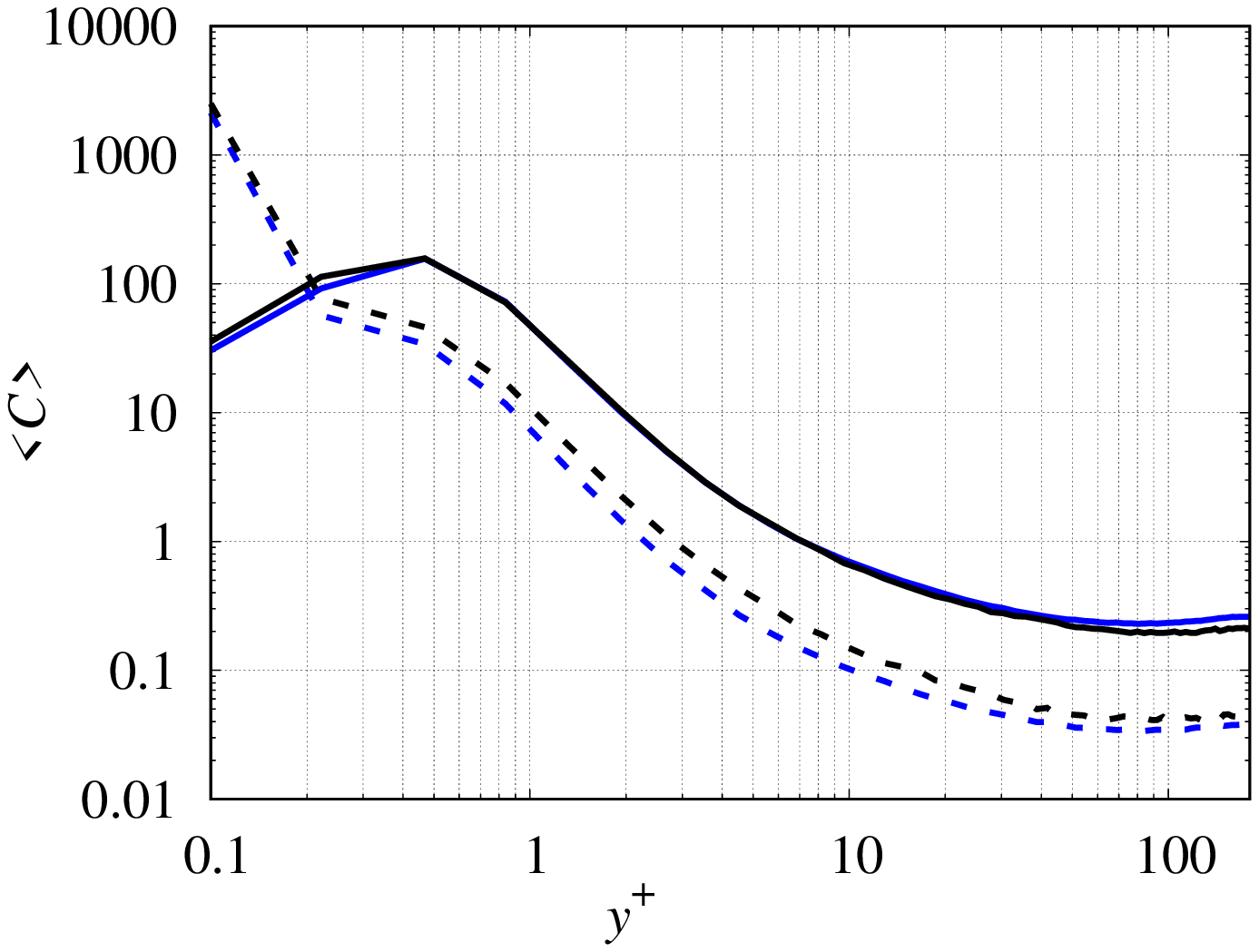}\label{fig:np4}}
\caption{Time-averaged particle concentration as a function of the distance from the wall.
For the duct, $y^+$ stretches from the bisector of one wall to the centreline of the duct.
}
\label{fig:np}
\end{figure*}

\begin{figure*}[tb]
\centering
\begin{tikzpicture}[thick]
\node [anchor=center] at (0,0)   {\includegraphics[trim=0mm 0mm 0mm 0mm,clip=true,height=.23\textwidth]{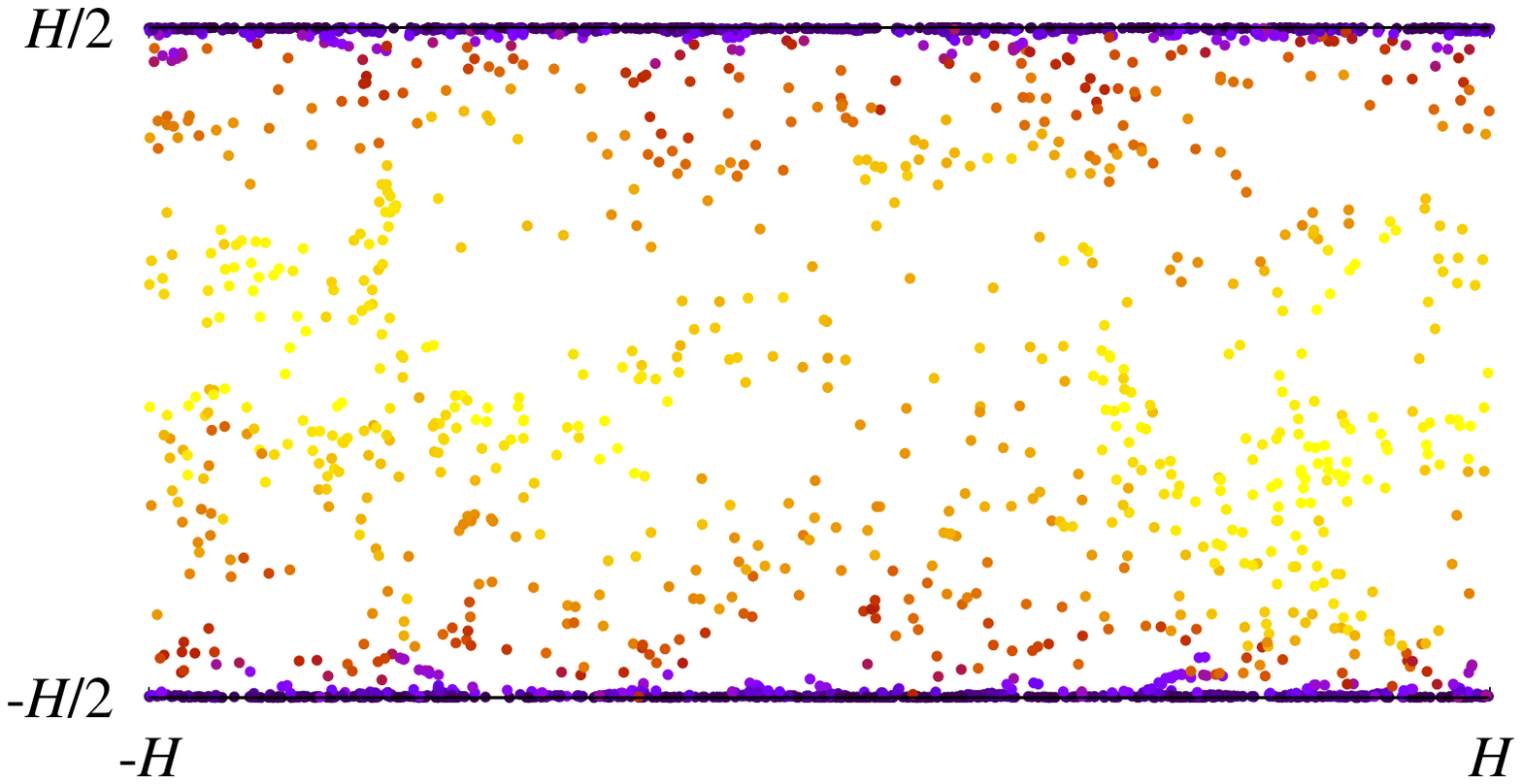}};
\node [anchor=center] at (5.7,0) {\includegraphics[trim=0mm 0mm 0mm 0mm,clip=true,height=.23\textwidth]{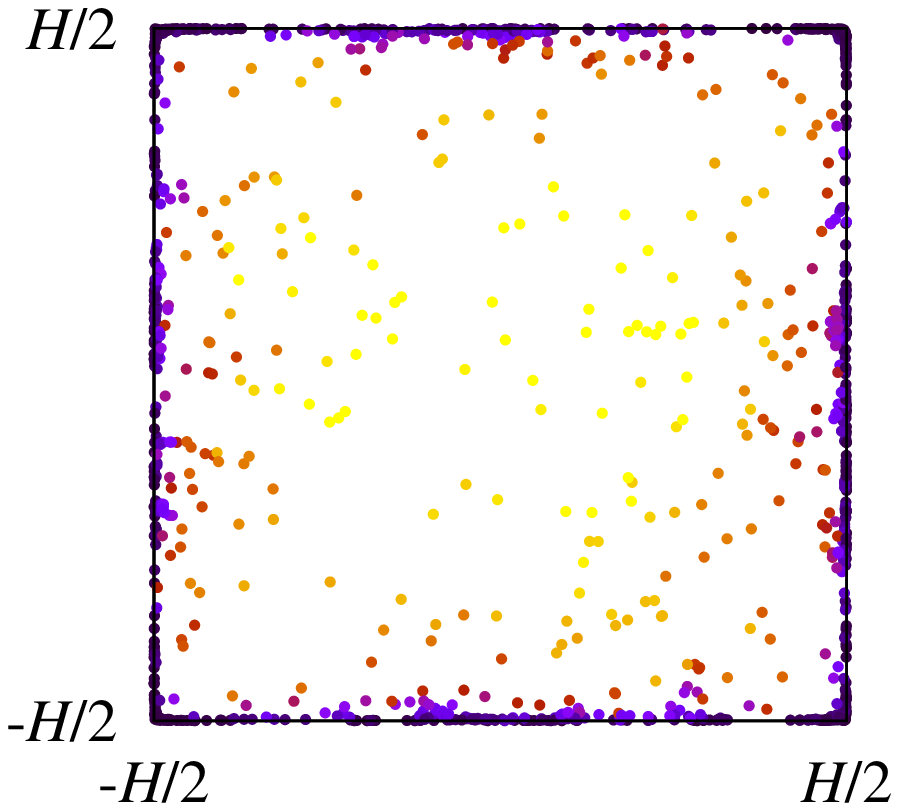}};
\node [anchor=center] at (10,0)  {\includegraphics[trim=0mm 0mm 0mm 0mm,clip=true,height=.23\textwidth]{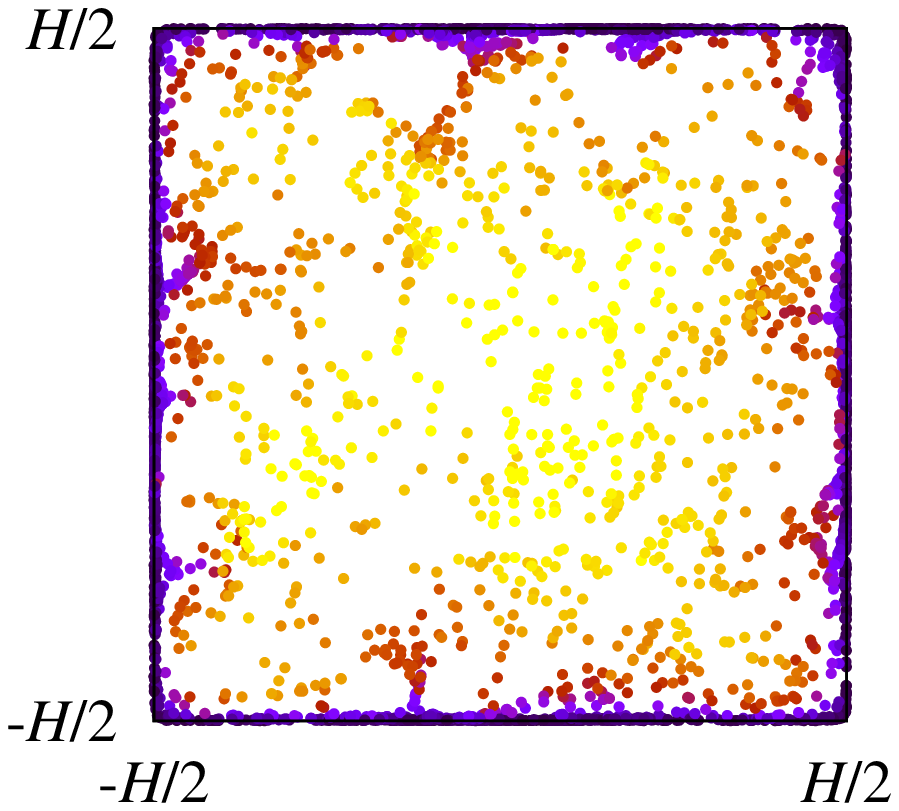}};
\node [anchor=center] at (0,-4)  {\includegraphics[trim=0mm 0mm 0mm 0mm,clip=true,height=.23\textwidth]{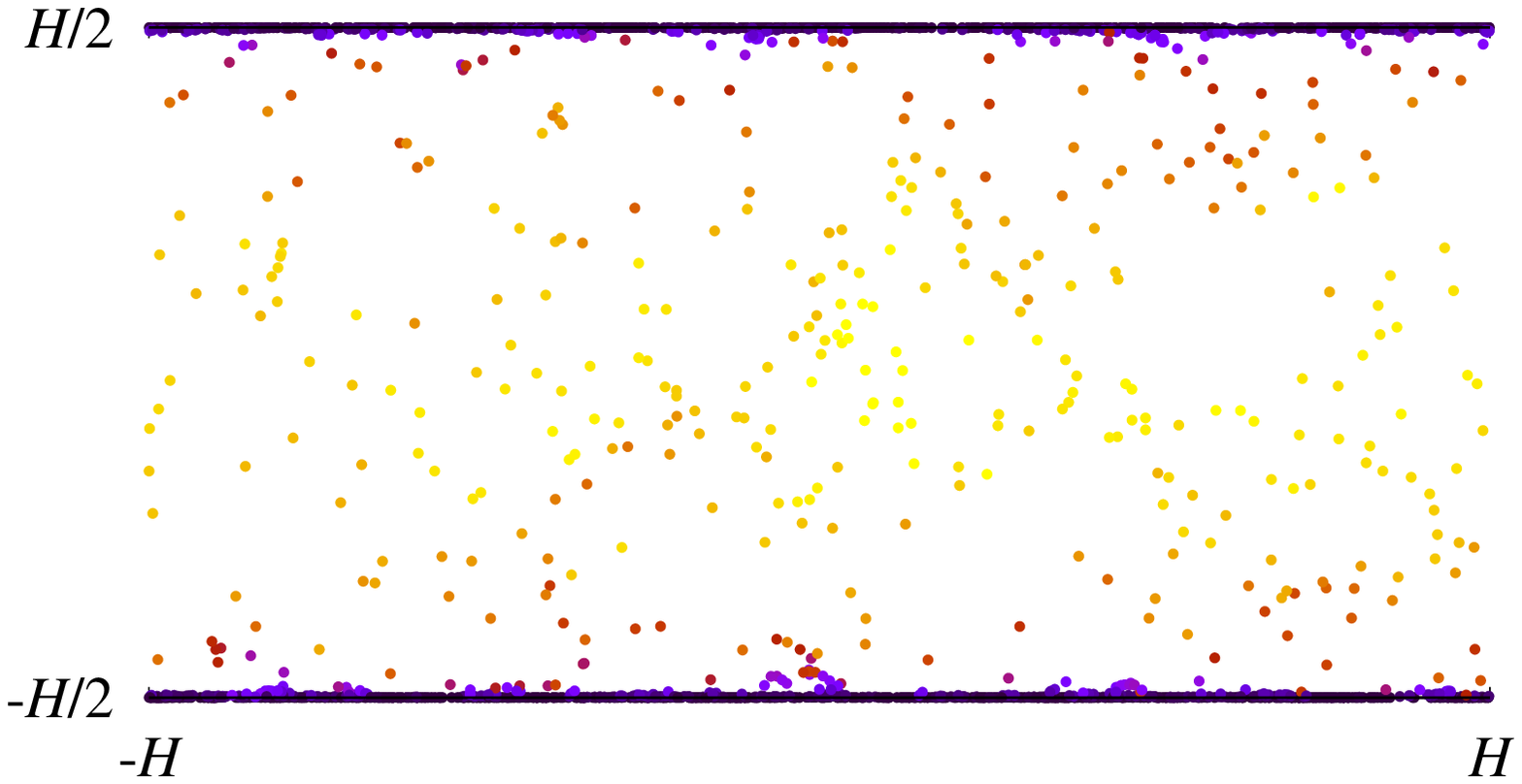}};
\node [anchor=center] at (5.7,-4){\includegraphics[trim=0mm 0mm 0mm 0mm,clip=true,height=.23\textwidth]{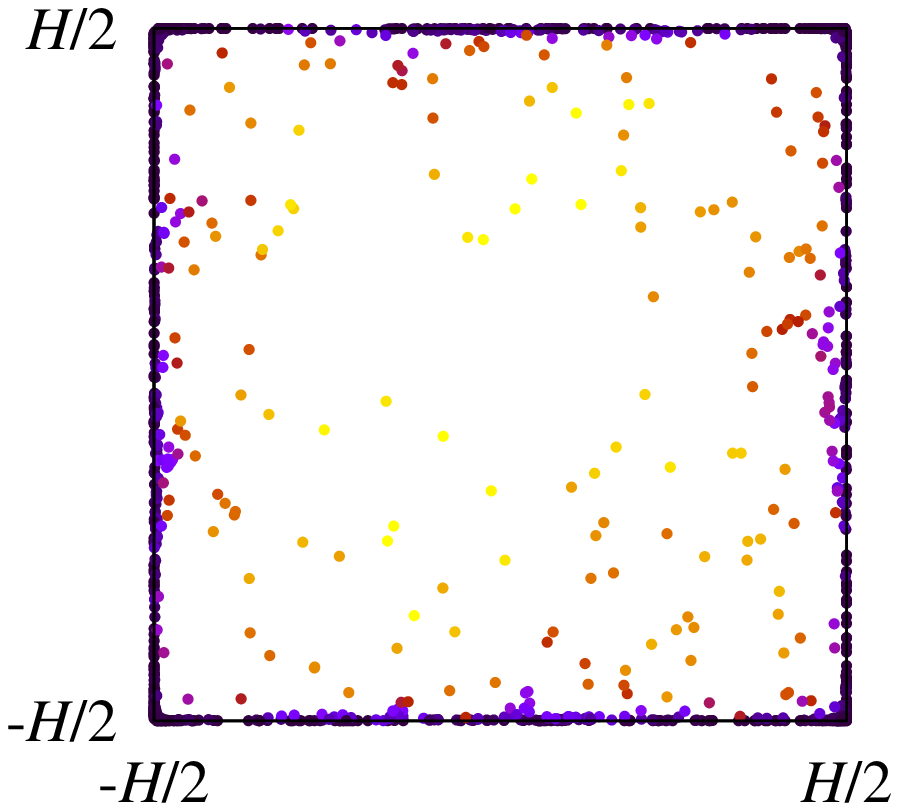}};
\node [anchor=center] at (10,-4) {\includegraphics[trim=0mm 0mm 0mm 0mm,clip=true,height=.23\textwidth]{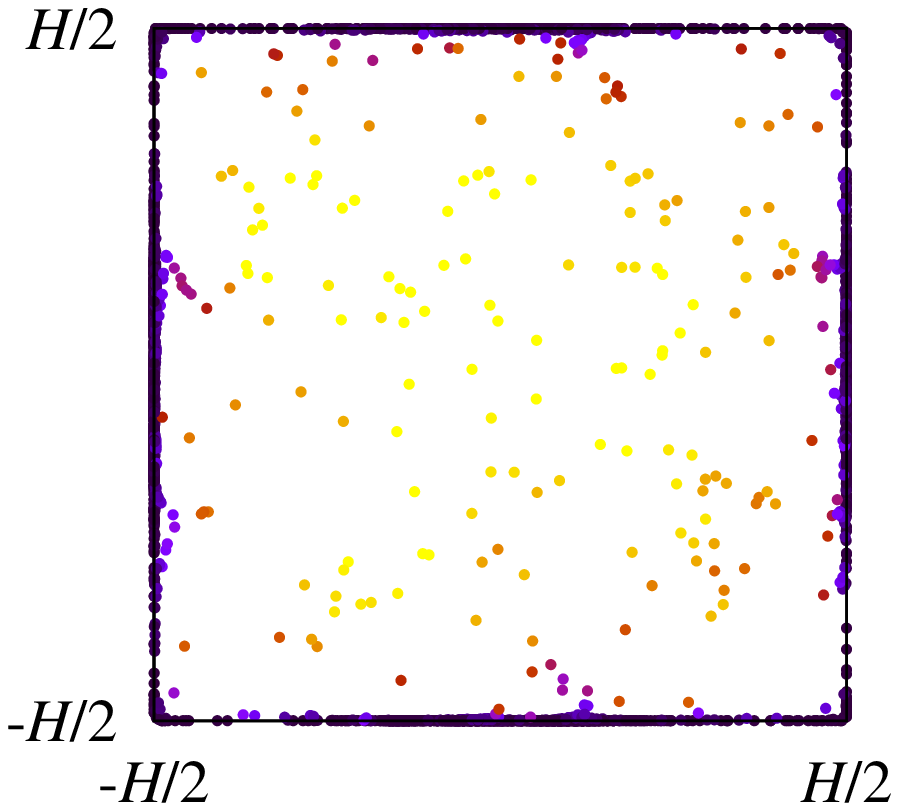}};
\node [anchor=center] at (.5,-9.5) {\includegraphics[trim=40mm 0mm 40mm 0mm,clip=true,height=.40\textwidth]{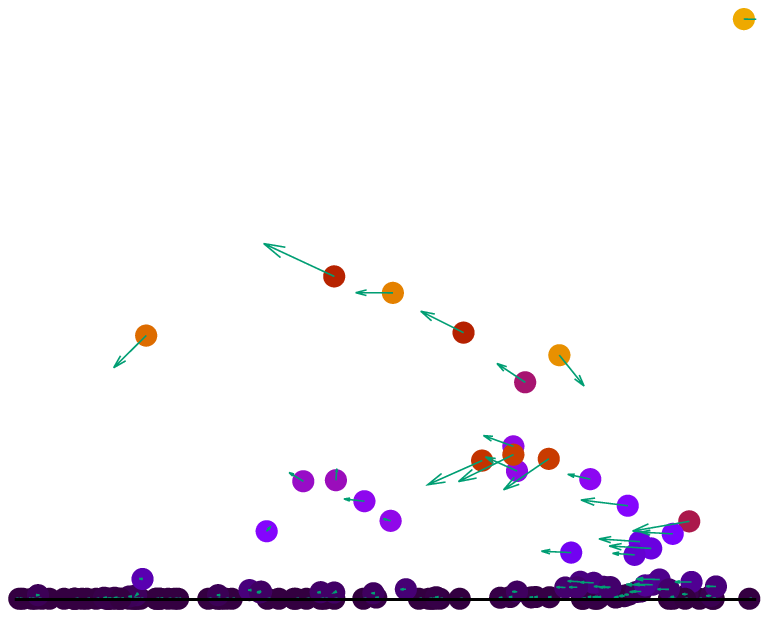}};
\node [anchor=center] at (8,-9.5)  {\includegraphics[trim=40mm 0mm 40mm 0mm,clip=true,height=.40\textwidth]{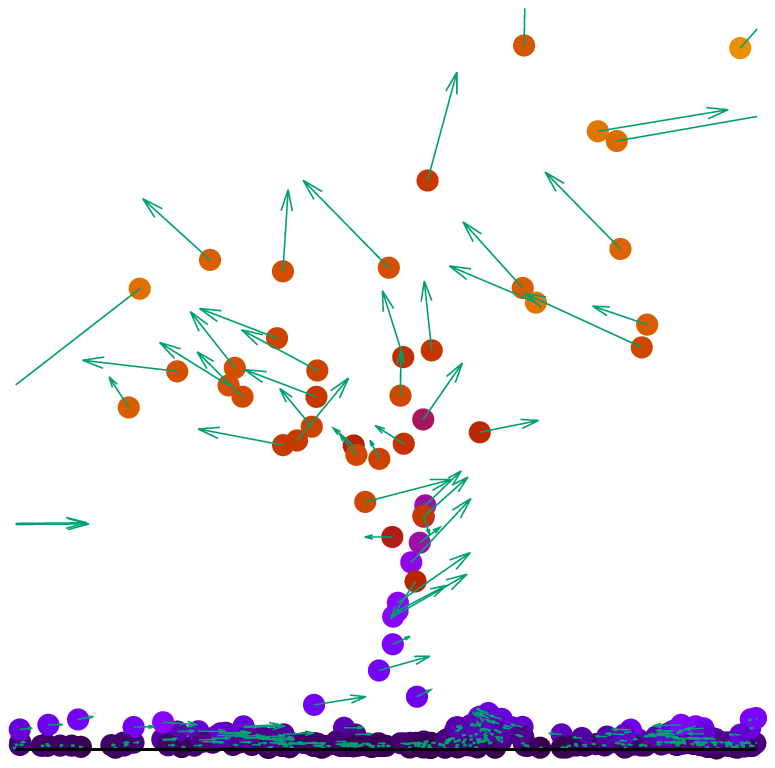}};
\draw [black,fill=blue,fill opacity=.1] (9.9,-1.2) circle [radius=.35];
\draw [black,opacity=.5,ultra thick,->] (9.4,-1.25) to [out=180,in=90] (8,-6.5);
\draw [black,fill=blue,fill opacity=.1] (0.2,-5.2) circle [radius=.35];
\draw [black,opacity=.5,ultra thick,->] (.2,-5.9) -- (.2,-7);
\node [anchor=center,align=center,rotate=90] at (-4,0) {uncharged};
\draw [black,opacity=.5,ultra thick,->] (-3.7,0) -- (-3.2,0);
\node [anchor=center,align=center,rotate=90] at (-4,-4) {charged};
\draw [black,opacity=.5,ultra thick,->] (-3.7,-4) -- (-3.2,-4);
\node [anchor=center,align=center] at (0,2.8) {Case 2, channel flow};
\draw [black,opacity=.5,ultra thick,->] (0,2.5) -- (0,2.0);
\node [anchor=center,align=center] at (6.0,2.8) {Case 2, duct flow};
\draw [black,opacity=.5,ultra thick,->] (6.0,2.5) -- (6.0,2.0);
\node [anchor=center,align=center] at (10.3,2.8) {Case 4, duct flow};
\draw [black,opacity=.5,ultra thick,->] (10.3,2.5) -- (10.3,2.0);
\end{tikzpicture}
\caption{Instantaneous particle positions of the cases~2 ($S\!t=32$, $S\!t_\mathrm{el}=0.004$) and 4 ($S\!t=8$, $S\!t_\mathrm{el}=0.004$).
The snapshots are recorded during the statistically stationary state of the simulation.
The view perspective is in streamwise direction.
The particle colors indicate their streamwise velocity, where yellow is the maximum of $u^+_\mathrm{p}=20$.
For clarity, the particles are enlarged and only 1/32 of the total channel/duct length is visualized.
}
\label{fig:upinst1}
\end{figure*}

\begin{figure*}[p]
\centering
\vspace{-5mm}
\subfigure[Case 1: $S\!t=8$, $S\!t_\mathrm{el}=0.002$] {\raisebox{-6mm}{\includegraphics[trim=0mm 0mm 0mm 0mm,clip=true,width=0.44\textwidth]{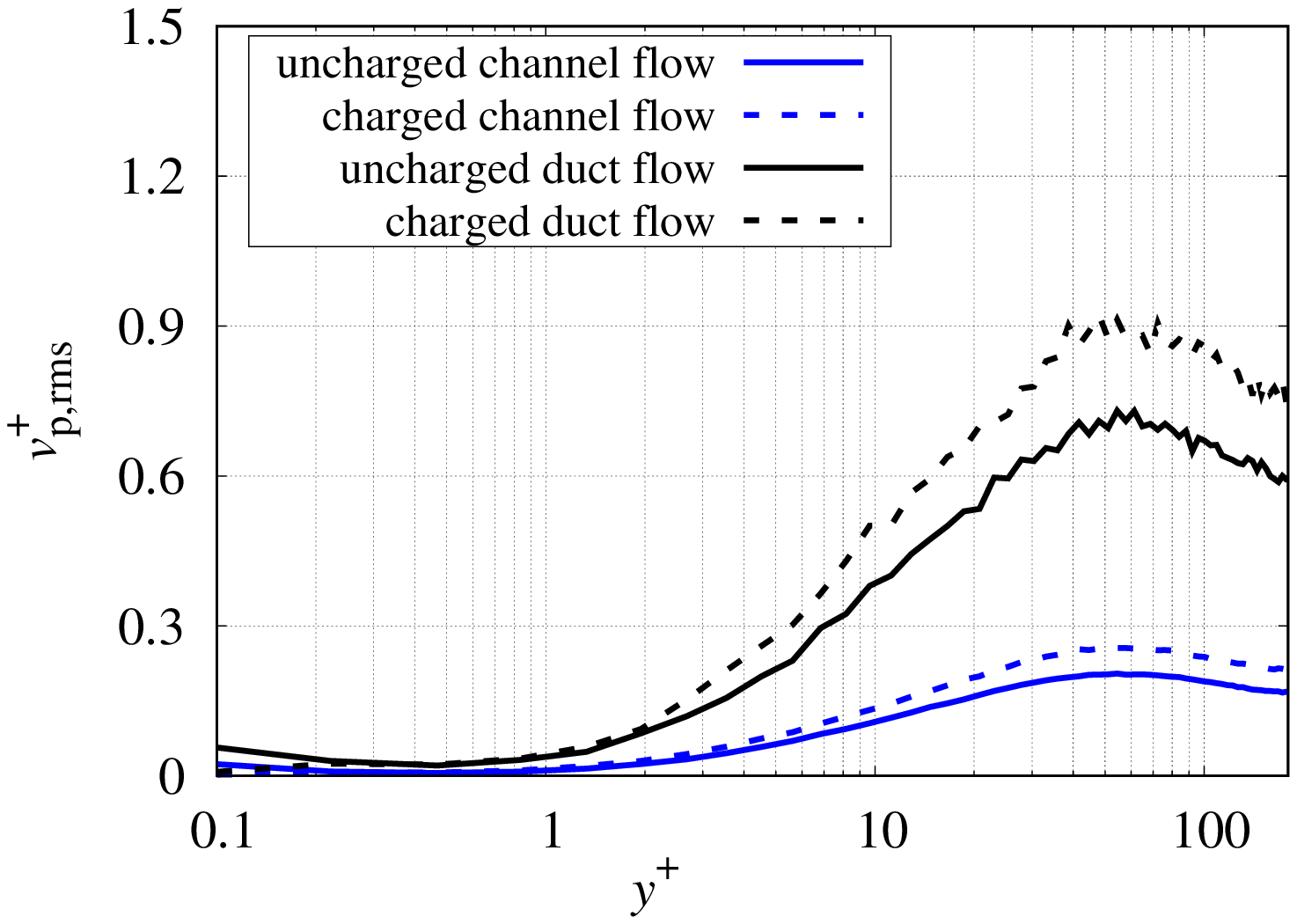}}\label{fig:vrms1}}\qquad
\subfigure[Case 1: $S\!t=8$, $S\!t_\mathrm{el}=0.002$] {\raisebox{-6mm}{\includegraphics[trim=0mm 0mm 0mm 0mm,clip=true,width=0.44\textwidth]{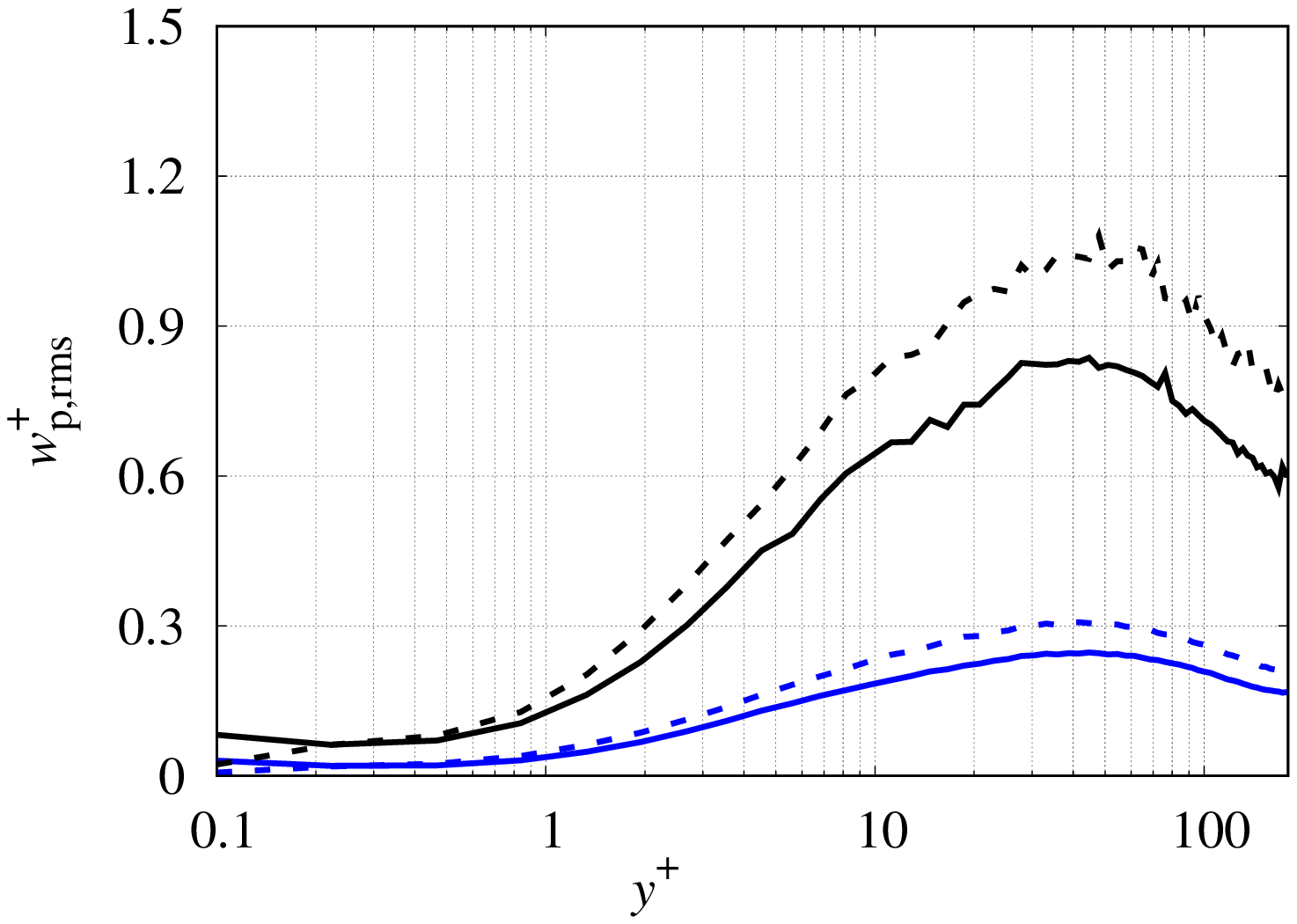}}\label{fig:wrms1}}\\
\subfigure[Case 2: $S\!t=32$, $S\!t_\mathrm{el}=0.004$]{\raisebox{-6mm}{\includegraphics[trim=0mm 0mm 0mm 0mm,clip=true,width=0.44\textwidth]{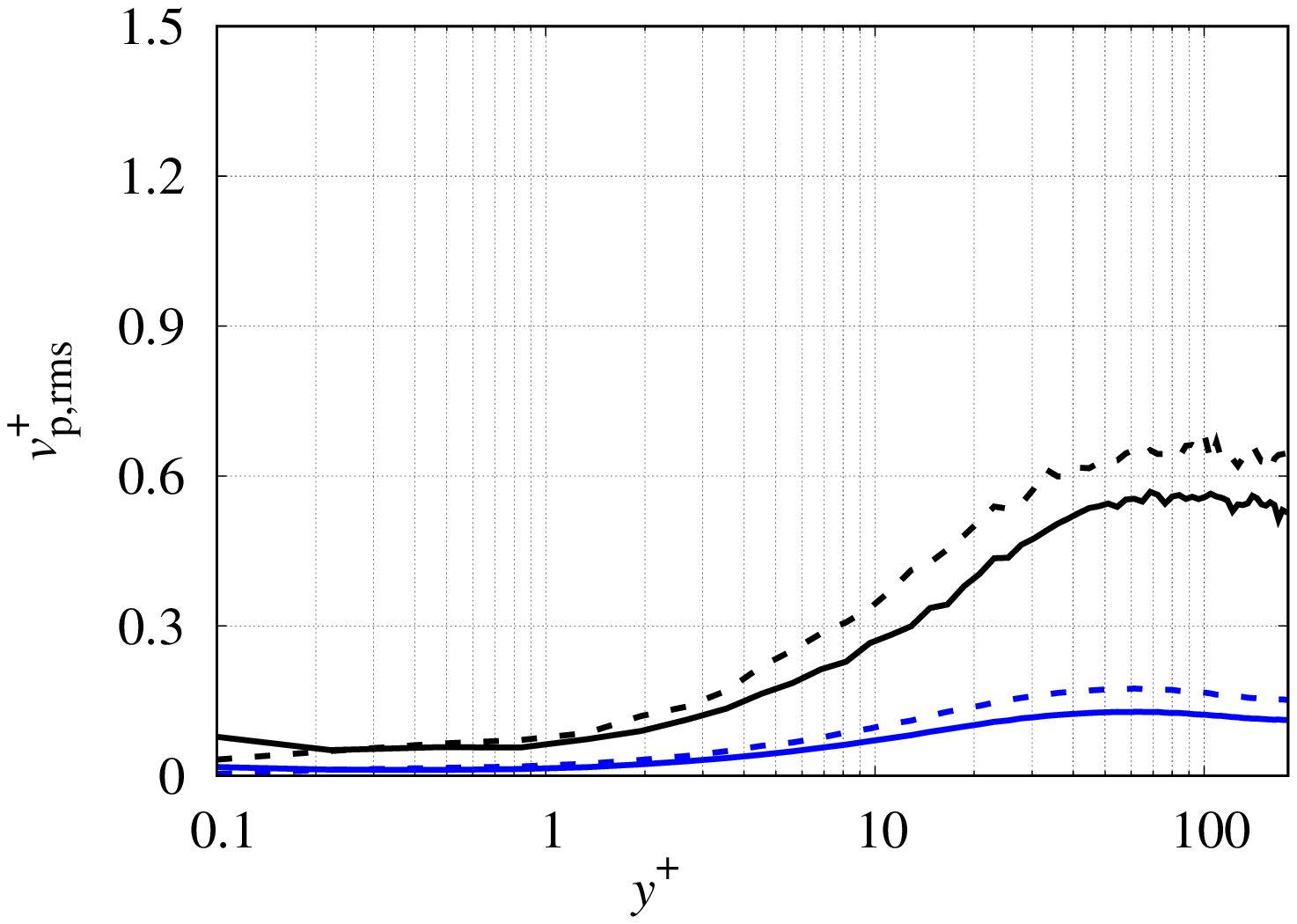}}\label{fig:vrms2}}\qquad
\subfigure[Case 2: $S\!t=32$, $S\!t_\mathrm{el}=0.004$]{\raisebox{-6mm}{\includegraphics[trim=0mm 0mm 0mm 0mm,clip=true,width=0.44\textwidth]{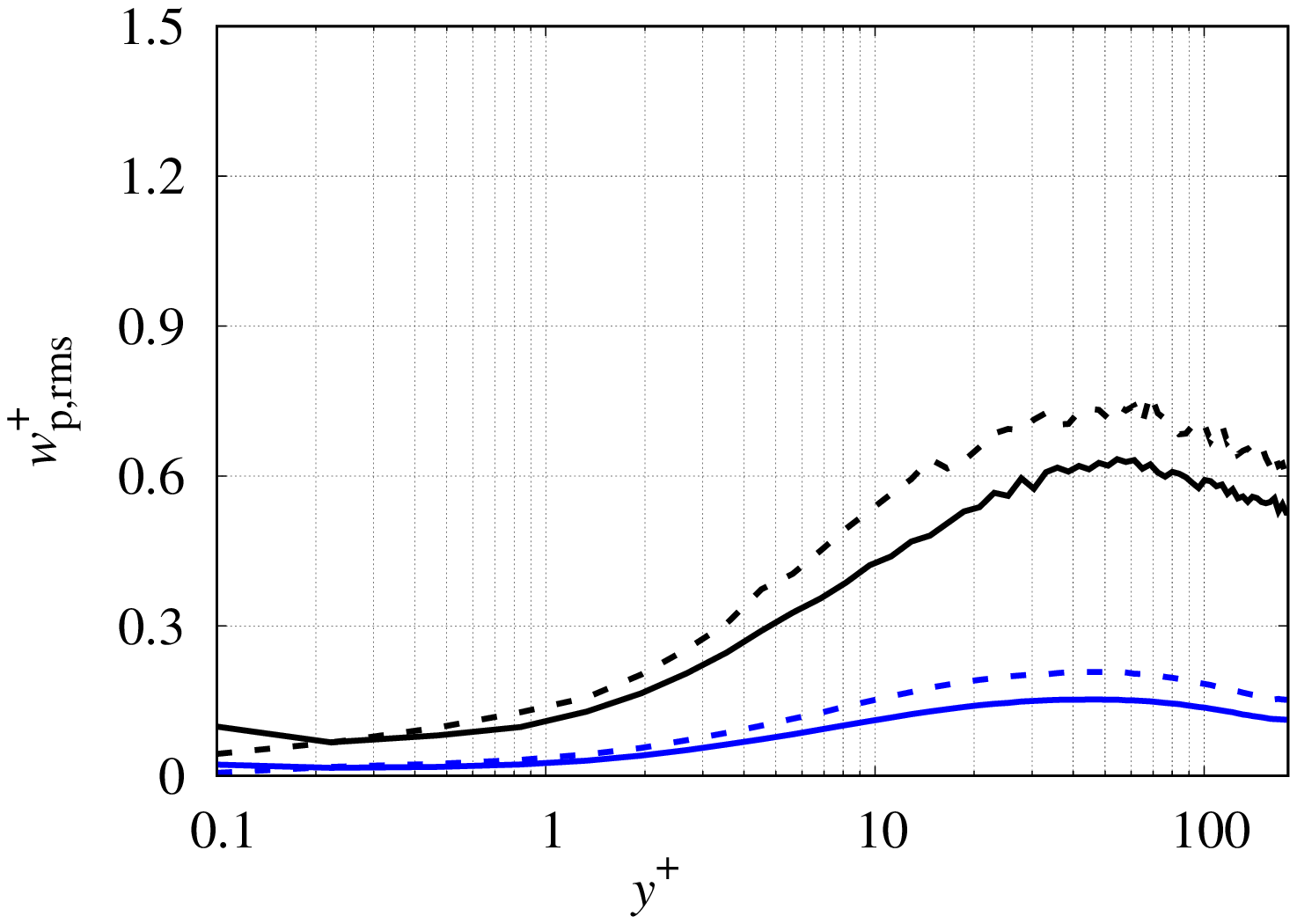}}\label{fig:wrms2}}\\
\subfigure[Case 3: $S\!t=8$, $S\!t_\mathrm{el}=0.001$] {\raisebox{-6mm}{\includegraphics[trim=0mm 0mm 0mm 0mm,clip=true,width=0.44\textwidth]{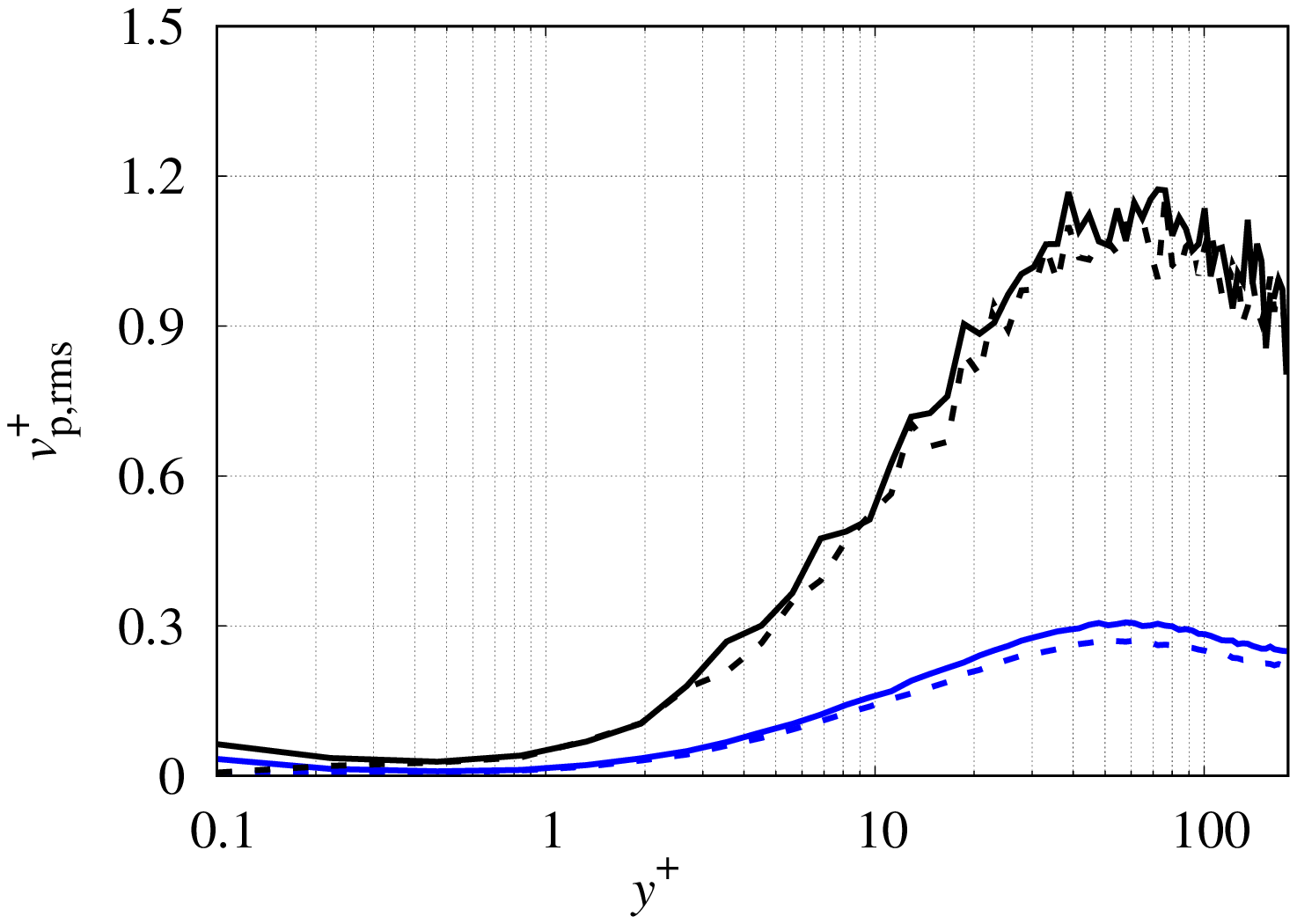}}\label{fig:vrms3}}\qquad
\subfigure[Case 3: $S\!t=8$, $S\!t_\mathrm{el}=0.001$] {\raisebox{-6mm}{\includegraphics[trim=0mm 0mm 0mm 0mm,clip=true,width=0.44\textwidth]{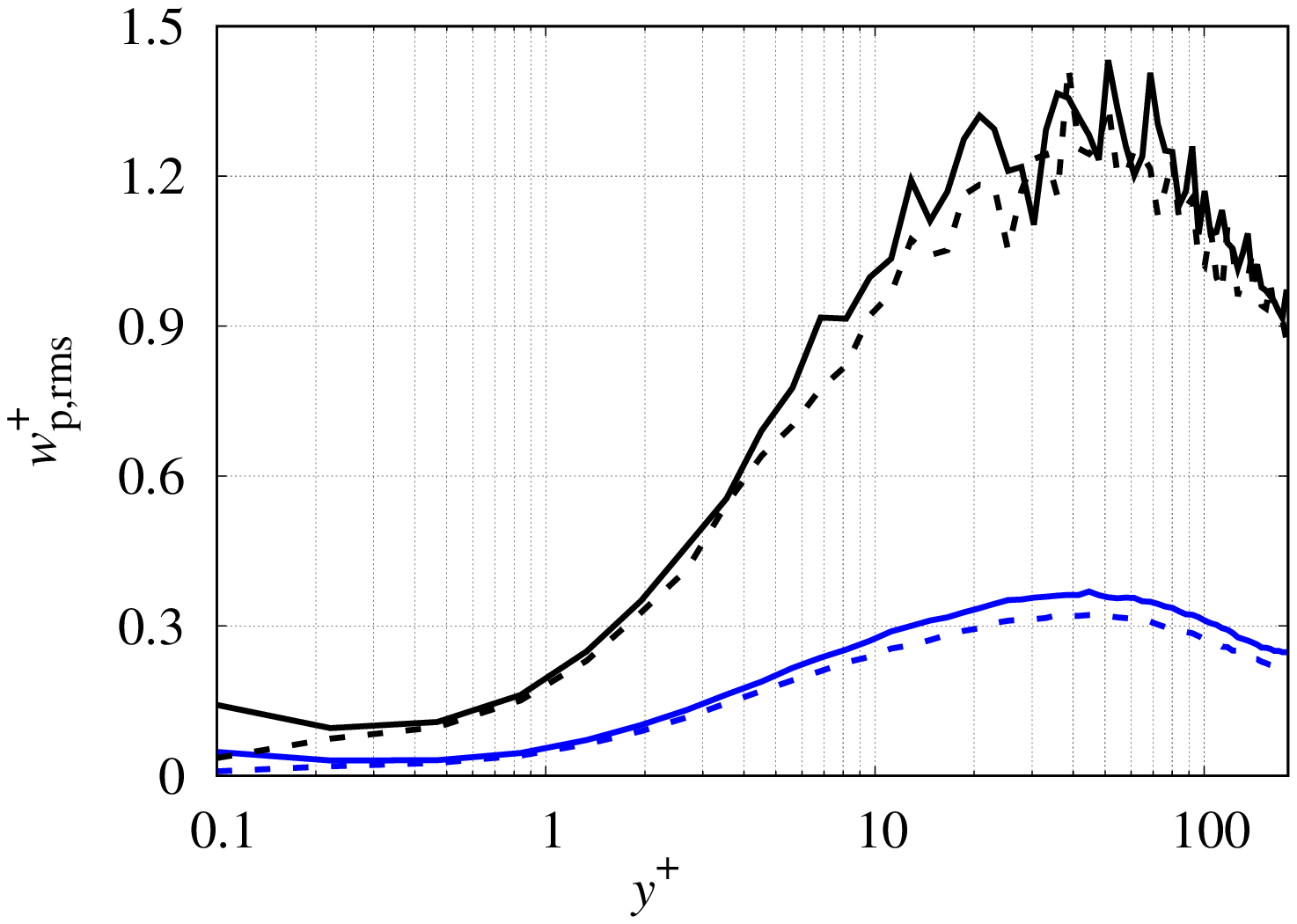}}\label{fig:wrms3}}\\
\subfigure[Case 4: $S\!t=8$, $S\!t_\mathrm{el}=0.004$] {\raisebox{-6mm}{\includegraphics[trim=0mm 0mm 0mm 0mm,clip=true,width=0.44\textwidth]{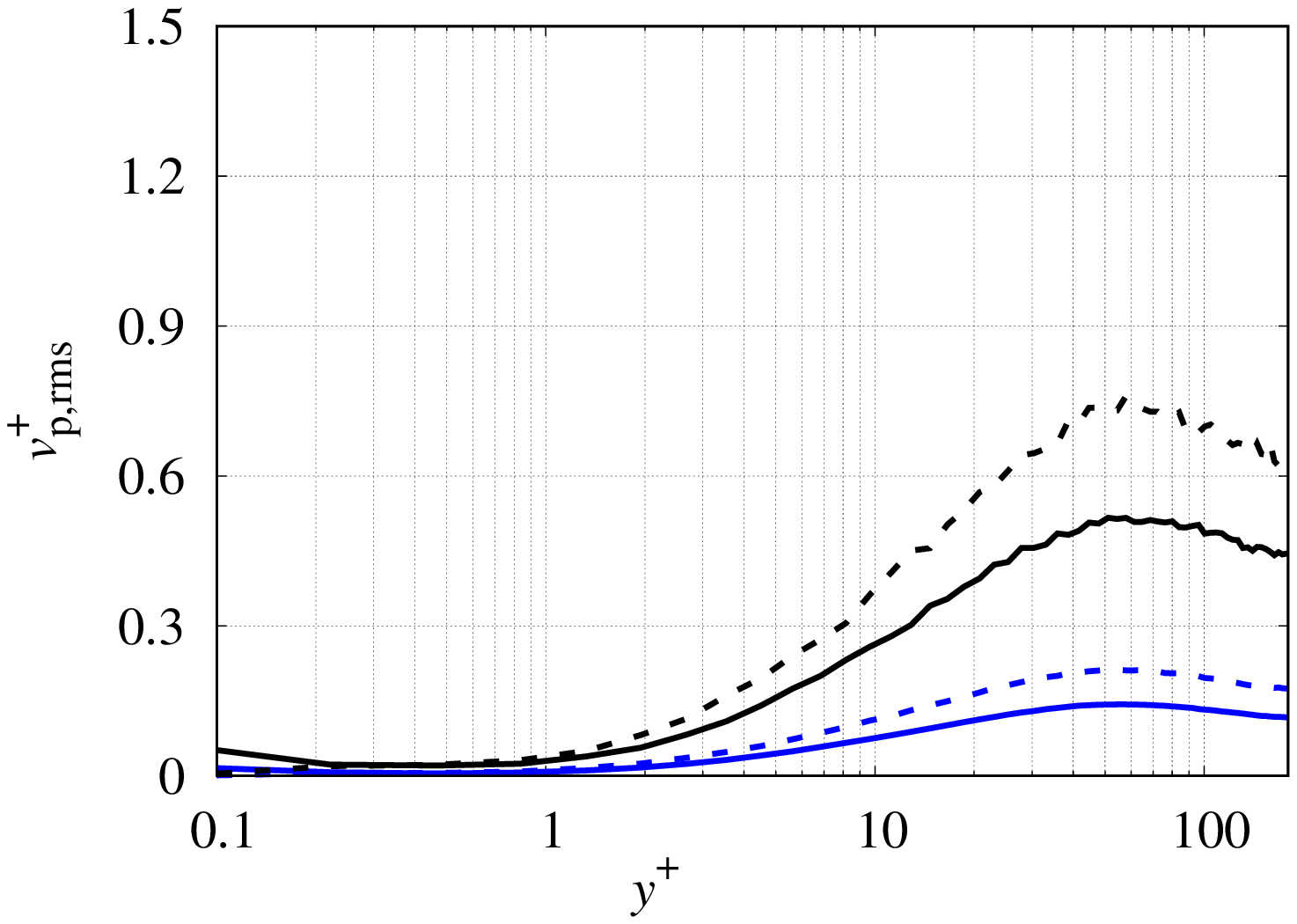}}\label{fig:vrms4}}\qquad
\subfigure[Case 4: $S\!t=8$, $S\!t_\mathrm{el}=0.004$] {\raisebox{-6mm}{\includegraphics[trim=0mm 0mm 0mm 0mm,clip=true,width=0.44\textwidth]{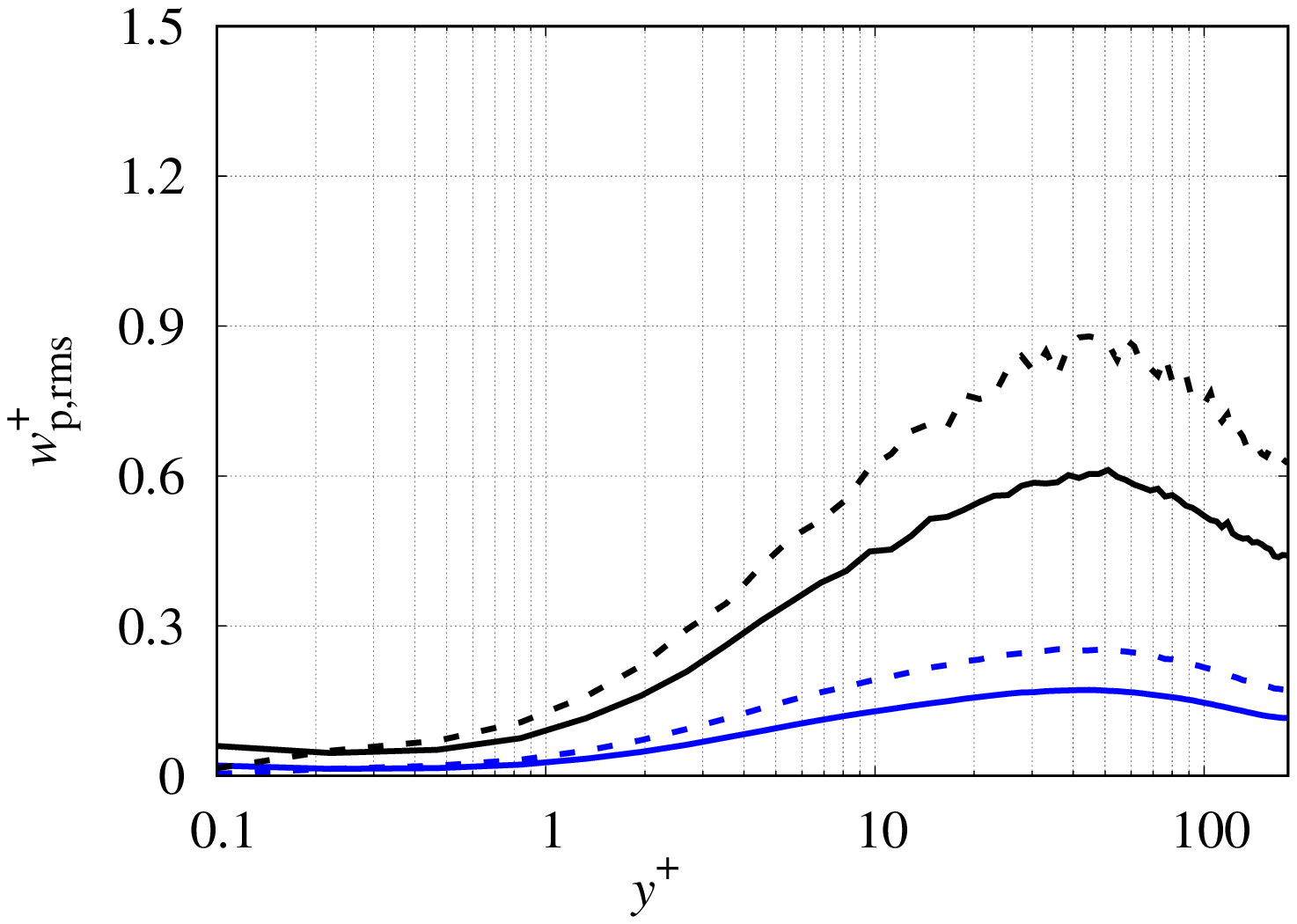}}\label{fig:wrms4}}
\caption{Fluctuations of the cross-sectional particle velocities as a function of the distance from the wall.
For the duct, $y^+$ stretches from the bisector of one wall to the centreline.
}
\label{fig:vwrms}
\end{figure*}

\begin{figure*}[tb]
\centering
\subfigure[]{\includegraphics[trim=0mm 6mm 0mm 0mm,clip=true,width=0.47\textwidth]{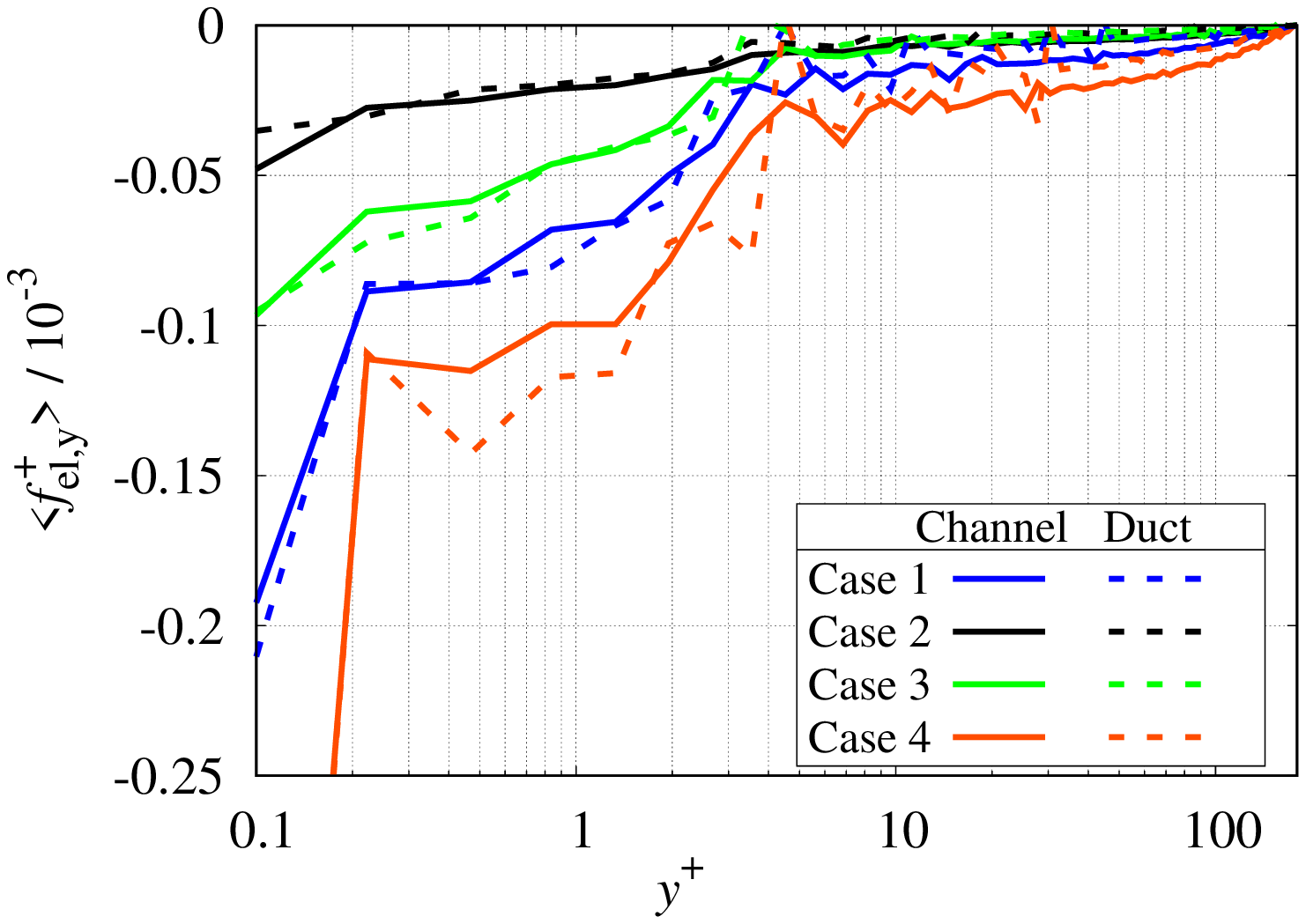}\label{fig:fyel}}\qquad
\subfigure[]{\includegraphics[trim=0mm 6mm 0mm 0mm,clip=true,width=0.47\textwidth]{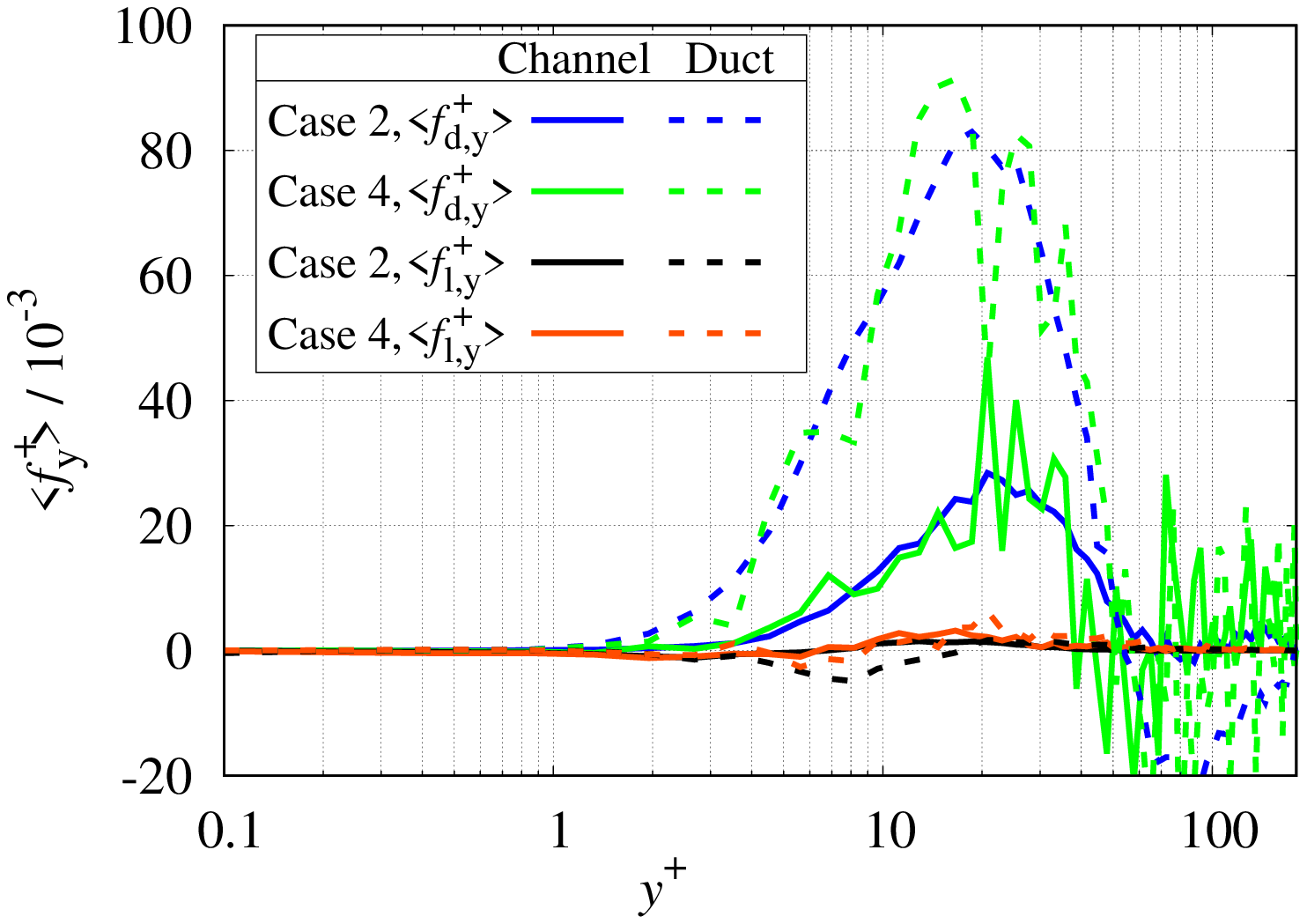}\label{fig:fy24}}\\
\caption{Time-averaged electrostatic, drag, and lift forces acting on the particles as a function of the distance from the wall.
Only charged-particle simulations are plotted.
For the duct, $y^+$ stretches from the bisector of one wall to the centreline of the duct.
\linebreak
(Case 1: $S\!t=8$, $S\!t_\mathrm{el}=0.002$;
Case 2: $S\!t=32$, $S\!t_\mathrm{el}=0.004$;
Case 3: $S\!t=8$, $S\!t_\mathrm{el}=0.001$;
Case 4: $S\!t=8$, $S\!t_\mathrm{el}=0.004$)
}
\label{fig:fy}
\end{figure*}

\begin{figure*}[tb]
\centering
\subfigure[Case 2: $S\!t=32$, $S\!t_\mathrm{el}=0.004$]{\includegraphics[trim=0mm 6mm 0mm 0mm,clip=true,width=0.47\textwidth]{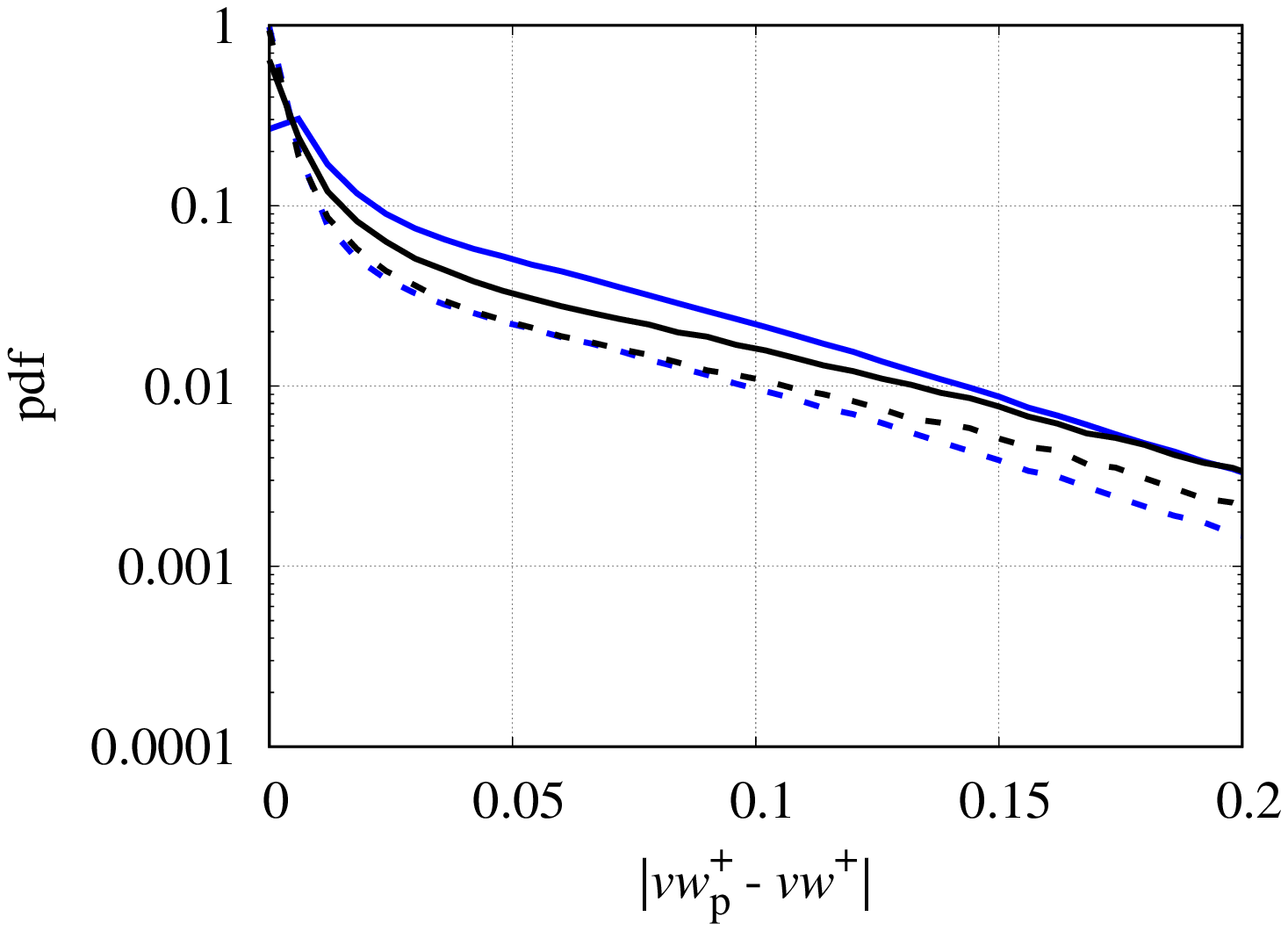}\label{fig:pdf-slip-2}}\qquad
\subfigure[Case 4: $S\!t=8$, $S\!t_\mathrm{el}=0.004$] {\includegraphics[trim=0mm 6mm 0mm 0mm,clip=true,width=0.47\textwidth]{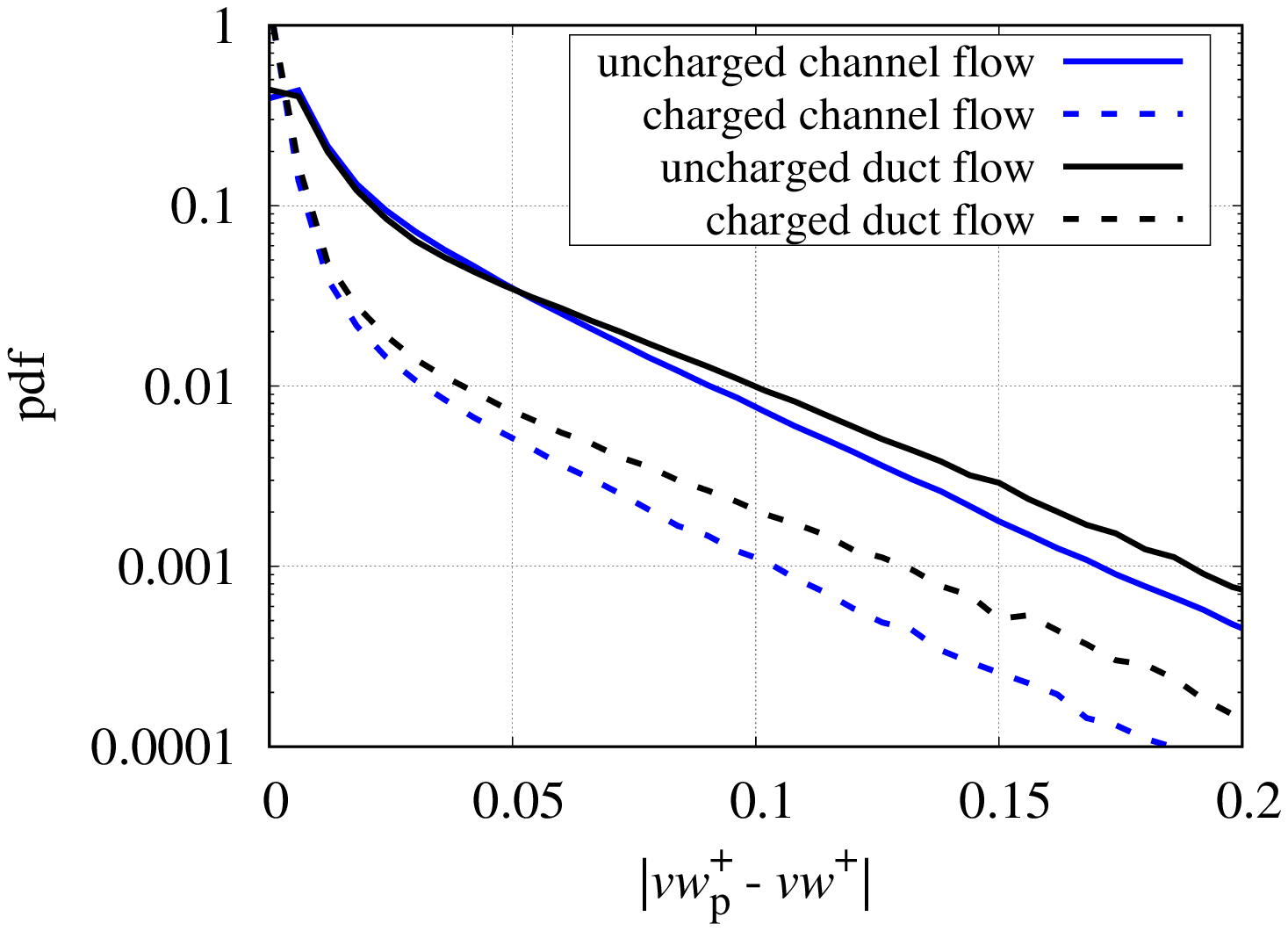}\label{fig:pdf-slip-4}}\\
\caption{Time-averaged streamwise particle velocities as a function of the distance from the wall.
For the duct, $y^+$ stretches from the bisector of one wall to the centreline of the duct.
}
\label{fig:pdf-slip}
\end{figure*}

The viscous sublayer extends five viscous units from a wall in the direction of the bulk and represents the region of the flow where the boundary layer is laminar.
Thus, $F_1$ quantifies the accumulation of particles in those regions free from turbulent fluctuations.
Since the particles are initially distributed homogeneously, at $t^+=0$, all $F_1$ equal the volume occupied by the first wall unit divided by the total volume of the domain, which is 0.0055 for the channel and 0.011 for the duct flows.
For the subsequent statistical analyses in this paper (figures~\ref{fig:np} to~\ref{fig:pdf-slip}, we ran all cases during the stationary phase to collect sufficient data.
The simulations stopped when the statistical parameters converged.
Hence, some curves do not extend the complete duration depicted in the graphs, and for some cases the figures do not cover the total simulation time.

At the beginning of the simulations, the aerodynamic forces tend to push particles from their initial positions toward the channel or duct walls due to the turbopheretic drift.
Additionally, if charges of equal polarity are assigned to the particles, repulsive electrostatic forces act in the same direction, from the flow's bulk to its periphery.
Thus, the particles convect further toward the walls if they are under the influence of electrostatic forces than only through turbopheretic forces.
Because the charged particles travel a longer distance from their initial to their preferential position, the concentration of the uncharged powder converges faster than the corresponding charged one.

Regarding uncharged flows, in case~2 $F_1$ adopts a lower value ($\approx 0.40$) than in the other cases ($\approx 0.55$), meaning the particles distribute more homogeneously.
Since case~2 is of the highest $S\!t$ number, the turbulent gas motion affects the particles the least.
In the other cases of lower $S\!t$, the particles disperse stronger, and $F_1$ becomes higher.
Each case's peak value of $F_1$ depends on the total particle number density.
The higher $\rho_N$, the more inter-particle collisions occur.
Collisions diffuse particles; they bounce more to less populated areas where they experience fewer collisions.
Therefore, the final values of $F_1$ are the highest for case~3 (lowest $\rho_N$) and descend slightly for cases~1 and~4.
Nevertheless, the influence of $\rho_N$ on $F_1$ is less than 2\%.

Strikingly stronger is the effect of electrical charges added to the particles.
The electrostatic forces transport the particles from their initial position toward the walls and increase $F_1$.
Even though the particles are only weakly charged, $F_1$ increases for all cases by more than 50\%.
Comparing the three cases of an equal Stokes number (1, 3, and~4), the fraction of particles that take a position close to the wall scales with the inverse of $S\!t_\mathrm{el}$.
If inertia is low, the influence of electrostatic forces on preferential particle positions becomes dominant.

Further, according to figure~\ref{fig:convergence}, electrostatic charges affect channel flows more than duct flows.
In channel flows, a larger fraction of particles accumulate close to the walls compared to duct flows.
The lower increase in duct flows is caused by vortices in the cross-section \citep{Gro20b}.
These vortices induce a motion of particles from the center of the duct along the diagonals towards the corners.
From there, the particles migrate along the walls to the wall’s bisectors, where they are ejected back into the bulk of the flow.
On the other hand, a channel flow does not exhibit net fluxes in the wall-normal or spanwise direction.
Due to this vortical motion, a duct flow is better mixed than a channel flow even if their parameters are otherwise identical.
Thus, the vortices in the ducts counteract the turbopheretic and electrostatic forces, which both push the particles toward the walls.

In sum, figure~\ref{fig:convergence}, shows that electrostatic forces, even if the charges are very low, strongly influence the particle flow pattern.
Also, secondary flows affect the particles.

To analyze this behavior in more detail, the particle concentration profiles (normalized by $\rho_N$) over the wall-normal coordinate are plotted in figure~\ref{fig:np}.
The spatial coordinate ($y^+$) is given in terms of wall units.
For the duct, this spatial axis stretches from the bisector of one wall ($y^+=0$, $z^+=180$) to the centreline ($y^+=180$, $z^+=180$) since at this slice duct and channel flows are the most similar.
The profiles are time-averaged during the statistically stationary phase of each simulation.
The averaging duration was long enough, so the profiles no longer changeid for further sampling.

Comparing the concentration profiles of charged and uncharged particles in each case confirms the observation made above:
The curves for the charged duct flows in \mbox{figure~\ref{fig:np}} are closer to the uncharged ones than the charged channel flows are to their uncharged counterparts.
Thus, the charges affect the particle distribution in the channel more than in the duct flows, even if $S\!t$ and $S\!t_\mathrm{el}$ are the same.

Interestingly, the concentration peaks attach to the walls when charge is assigned.
This results from electrostatic forces in-between particles, which push the particles further toward the walls.
On the other hand, for all uncharged simulations, the peak forms either at $y^+=0.2$ ($S\!t=32$) or $y^+=0.5$ ($S\!t=8$).

The electrostatic forces are in sum the larger, the more particles are present, i.e., they scale with $S\!t_\mathrm{el}$.
For $S\!t=8$ and $S\!t_\mathrm{el}=0.001$, the peak at the wall reaches a concentration of 1000.
For the same $S\!t$ but $S\!t_\mathrm{el}=0.002$, the concentration peak becomes 1700, and for a further increase to $S\!t_\mathrm{el}=0.004$ the peak is 2300.

On the other hand inertial forces become more important, through an increase of $S\!t$.
Thus, the distributions of charged particles in case~3 are closer to the uncharged distributions than in the other cases.

The above discussion is further supported by the snapshots of the instantaneous particle positions in figure~\ref{fig:upinst1}.
The figure shows cases~2 and~4, which are the extremes of our experimental design;
case~2 is the least affected by electrostatic forces due to the high particle inertia, and case~4 is the most affected due to the high particle number density.
For both cases, one can observe that the particles in the uncharged flows form agglomerations and structures which are triggered by local turbulent eddies, a phenomenon that has been elaborated on previously \cite{Marc02}.
Repulsive electrostatic forces dissolve these clusters.
Also, electrostatic effects reduced the particle density in the center of the channel for all conditions.

The detail in figure~\ref{fig:upinst1} of the uncharged duct flow of case~4 visualizes the phenomenon that causes cross-sectional mixing in duct flows:
in-plane vortices frequently grab particles residing close to a wall.
The particles eject in plume-like structures back to the center of the flow.
The velocity vectors indicate the wall-normal particle acceleration.
Several of such plumes are visible in the duct's cross section.

When charge is assigned to the particles, the electrostatic forces hinder plumes.
Only a few, rather small ones, can be observed in the snapshot.
Similarly, for the duct flow of case~2, the electrostatic forces hinder the ejection of particles from the wall. 

The least plumes occur for the charged channel flow of case~2.
Since the channel flow forms no secondary flows, only weaker turbulent fluctuations drive particles in wall-normal direction.
Electrostatic forces even counter these weak wall-normal movements.
In the snapshot in figure~\ref{fig:upinst1}, only few plume-like structures emerge from the near-wall region.
The detail of the largest visible shows low wall-normal velocities of the particles in the plume.
Thus, the particles remain close to the wall and their concentration increases.

Figure~\ref{fig:vwrms} plots the rms of the particle velocities in wall-normal and spanwise direction.
In this and the following figures, the velocity components are normalized with $u_\tau$, which is indicated by the superscript '+'.
This data represents the rms of the total velocities and not only the fluctuations, i.e.,
\begin{equation}
v^+_\mathrm{p,rms} = \sqrt{\dfrac{1}{T_2 - T_1} \int_{T_1}^{T_2} \left(v^+_\mathrm{p}\right)^2 dt} \, ,
\end{equation}
where $T_1$ and $T_2$ are time instances during the statistically stationary phase of the flow.
The quantities $u^+_\mathrm{p,rms}$ and $w^+_\mathrm{p,rms}$ are derived equivalently.

For all $S\!t$ and $S\!t_\mathrm{el}$, the wall-normal ($v^+_\mathrm{p,rms}$) and perpendicular ($w^+_\mathrm{p,rms}$) mobility of the particles in the duct is much higher than in the channel flow.
This higher mobility is caused by the vortical particle motions induced by the duct's secondary flow of the carrier gas.
The heavier particles of $S\!t=32$ follow the gas phase less than those of lower $S\!t$;
thus, their level of turbulence-induced fluctuations is lower.

For both types of flows, the particle velocity fluctuations increase significantly once charge is assigned.
These additional fluctuations are due to repulsive forces in-between charges.
An uncharged particle flying in the wake of another particle tends to adapt to the velocity of the preceding one~\citep{Prahl07,Prahl07-2}.
On the contrary, close charged particles repulse each other.
Thus, they accelerate in all spatial directions independent of the local flow field.
The effect of the electric forces on particle dynamics scales with $S\!t_\mathrm{el}$;
the effect even diminishes for the lowest $S\!t_\mathrm{el}$ of 0.001.

Case~3 contains the least particles.
Therefore, their related statistics converge slowly and the velocity profiles in figures~\ref{fig:vrms3} and~\ref{fig:wrms3} scatter more than those of the other cases.

Figure~\ref{fig:fy} depicts the time-averaged electrostatic, drag, and lift forces acting on the particles as a function of the distance from the wall.
The forces are normalized by the particle's mass and the fluid's frictional quantities.

The electrostatic forces in figure~\ref{fig:fyel} are one order of magnitude lower than the lift forces (figure~\ref{fig:fy24}), which are again one order of magnitude lower than the drag forces (figure~\ref{fig:fy24}).
Recalling the above discussed influence of the electrostatic forces, their magnitude is surprisingly low.
Electrostatic forces, being two orders of magnitude lower than drag forces, increase the near-wall particle concentration by two orders of magnitude (figure~\ref{fig:np}).

The magnitude of electrostatic forces scales inversely with the square of the particle spacing (cf.~equation~(\ref{eq:fel})).
Consequently, the forces amplify close to the walls where the particle concentration is the highest.
Also, the electrostatic forces scale with $S\!t_\mathrm{el}$ and $S\!t$.
They are the strongest the higher $S\!t_\mathrm{el}$ and the lower $S\!t$ is.
On the other hand, secondary flows have a minor influence on $f_{\mathrm{el},y}$.

As observed in the previous figures, cases~2 and~4 envelope our experimental design;
case~2 being the weakest and case~4 the strongest influenced by charge.
For clarity, figure~\ref{fig:fy24} contains only of those two cases the drag and lift force distribution.
Also, those curves are the same for uncharged and charged simulations.
Therefore, only charged simulations are plotted.
Both, $f_{\mathrm{d},y}$ and $f_{\mathrm{l},y}$ are larger for the duct than the channel, which is due to the enhanced spanwise particle mobility in the duct (cf.~figure~\ref{fig:vwrms}).

Finally, we check whether electric charges decorrelate the fluid and particle velocity, as it does in homogeneous turbulence.
To this end, figure~\ref{fig:pdf-slip} plots for cases~2 and~4 the pdfs of the spanwise slip velocity, which reads
\begin{equation}
\left| vw^+_\mathrm{p} -vw^+ \right| = \left( \left( v^+_\mathrm{p}-v^+\right)^2 + \left(w^+_\mathrm{p}-w^+\right)^2\right)^{1/2} \, .
\end{equation}
According to these cases, for channel and duct flows, electric charges decrease the slip velocity.
In other words, electric charges correlate the fluid and particle velocities.

In channel flows, the probability that a particle of $S\!t=32$ has a slip velocity of 0.01 or more decreases by approximately if it carries charge.
For particles of $S\!t=8$, the probability decreases even by nearly a factor of~10.
Since secondary flows stabilize the powder flow, the effect on duct flows is less than on channel flows.

The slip velocity decreases because electrostatic forces push the particles toward the walls.
Then, the slow fluid close to the walls decelerates the particles until they obtain a similarly low velocity.
Thus, the slip velocity decreases.

To sum up, contrary to particles transported by homogeneous turbulence, particles subjected to heterogeneous, wall-bounded turbulence increase their slip velocity once charge is assigned.
Once again, relatively weak electrostatic forces and secondary flows have a striking influence on particle flow patterns.

\section{Conclusions}
The data presented in this paper demonstrates the strong influence even of weak electrostatic forces and secondary flows on powder pneumatic transport.
Our DNS allowed drawing detailed conclusions regarding small-scale flow mechanisms.
In particular, unipolar charges support the migration of particles towards the boundaries of the flow domain.
Thus, electrostatic forces enhance the effect of turbopheresis, increasing the particle concentration close to the walls.
Also, they counter the flow-driven ejection of particles from the near-wall to the central region of the flow.
Contrary to particles in homogeneous turbulence, particles in wall-bounded turbulence decrease their slip velocity.
Secondary flows in ducts stabilize the particle flow pattern, and their dynamics is less affected by electrostatics.
Given the importance of the local particle concentration and charge accumulation, these results are relevant for the safety and operational performance of manifold technical powder flow facilities.
Especially the future exploration of polydispersity, charge distributions, and bipolar charges, including the modeling of the charge-build-up depending on the transport parameters, promises a further understanding of the mechanisms underlying this type of flows.

\section*{Acknowledgments}

This project has received funding from the European Research Council (ERC) under the European Union’s Horizon 2020 research and innovation programme (grant agreement No. 947606 PowFEct).

\section*{Data Availability Statement}
Raw data were generated at the PTB High Performance Computing cluster.
Derived data supporting the findings of this study are available from the corresponding author upon reasonable request.

\bibliography{\string~/essentials/publications}



\end{document}